\documentclass[fleqn,usenatbib,onecolumn]{mnras}

\usepackage{newtxtext,newtxmath}

\usepackage[T1]{fontenc}
\usepackage{ae,aecompl}

\usepackage{graphicx}	
\usepackage{amsmath}	
\usepackage{amssymb}	
\usepackage{soul}

\newcommand{\Kepler}{{\sl Kepler}\ }

\newcommand{\be}{\begin{equation}}
\newcommand{\ee}{\end{equation}}
\newcommand{\dir}{\textnormal{d}}

\title[Circumbinary Instabilities]{Instabilities in Multi-Planet Circumbinary Systems}

\author[Sutherland \& Kratter]{
Adam~P.~Sutherland,$^{1}$\thanks{E-mail: adamsutherland@email.arizona.edu}
Kaitlin~M.~Kratter,$^{1}$
\\
$^{1}$Steward Observatory, University of Arizona, 933 N. Cherry Ave., Tucson, AZ 85721, USA\\
}

\date{Accepted XXX. Received YYY; in original form ZZZ}

\pubyear{2019}

\begin{document}
\label{firstpage}
\pagerange{\pageref{firstpage}--\pageref{lastpage}}
\maketitle

\begin{abstract}
The majority of the discovered transiting circumbinary planets are located very near the innermost stable orbits permitted, raising questions about the origins of planets in such perturbed environments. Most favored formation scenarios invoke formation at larger distances and subsequent migration to their current locations. Disk-driven planet migration in multi-planet systems is likely to trap planets in mean motion resonances and drive planets inward into regions of larger dynamical perturbations from the binary. We demonstrate how planet-planet resonances can interact with the binary through secular forcing and mean-motion resonances, driving chaos in the system. We show how this chaos will shape the architecture of circumbinary systems, with specific applications to Kepler 47 and the Pluto-Charon system, limiting maximum possible stable eccentricities and indicating what resonances are likely to exist. We are also able to constrain the minimum migration rates of resonant circumbinary planets.
\end{abstract}

\begin{keywords}
planetary systems; planets and satellites: dynamical evolution and stability
\end{keywords}



\section{Introduction}

\begin{figure}
\centering
\includegraphics[scale=.5]{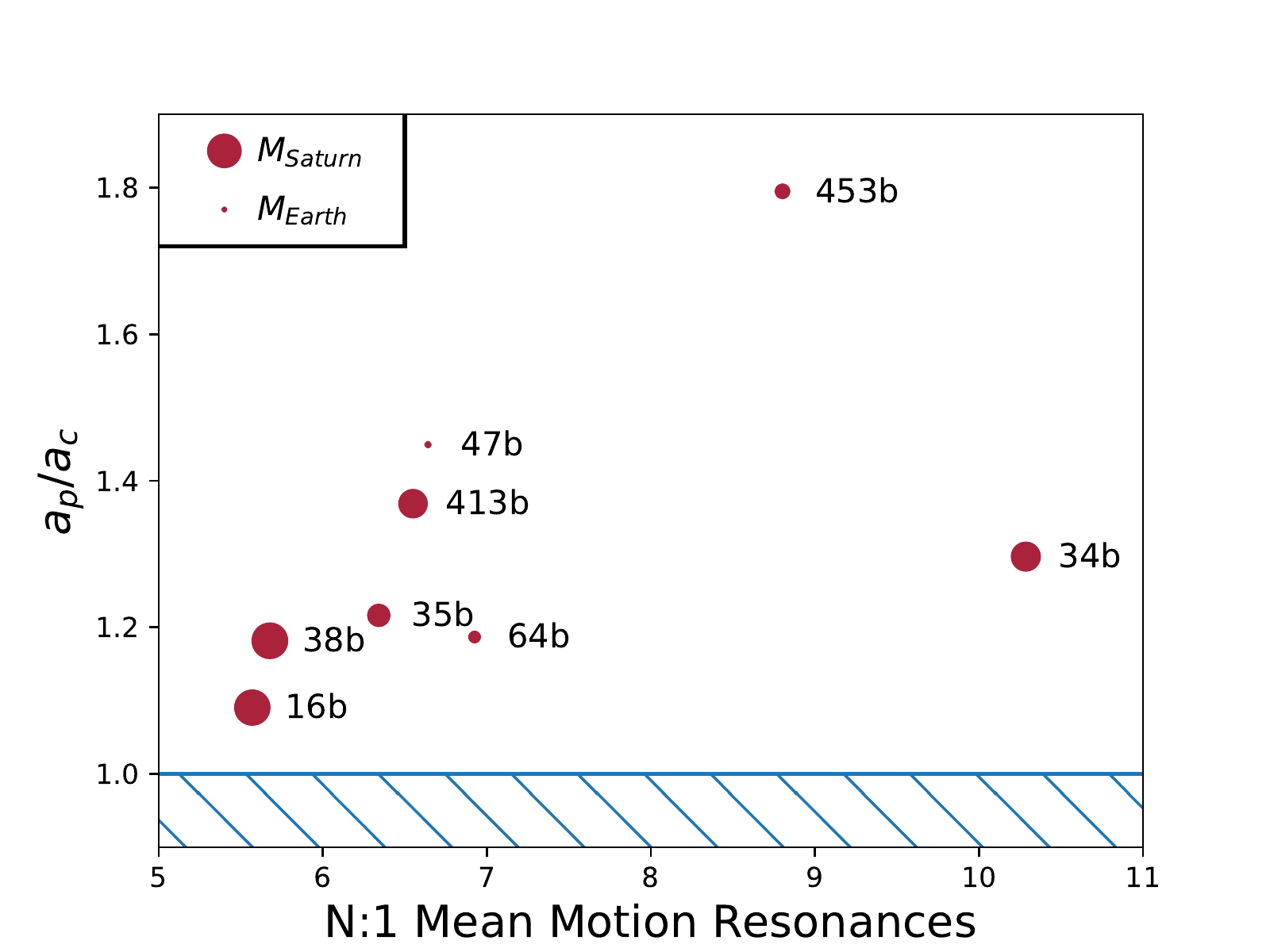} 
\caption{  
Ratio of semi-major axis to critical stable semi-major axis vs location of \Kepler circumbinary planets in terms of mean motion resonances with the binary. The size of each point corresponds to the size of the planet.}
\label{fig:n1_crit}
\end{figure}

The discovery of circumbinary planets has motivated new theories of planet formation and evolution in the presence of large dynamical perturbations. The majority of the well characterized circumbinary planets discovered via the \Kepler mission have orbits close to their host stars \citep{pile_up} near the minimum distance a circumbinary planet can remain on a stable orbit over long timescales \citep{HW1999}. The stability limit is due to overlapping mean motion resonances with the binary and is an efficient method of ejecting circumbinary planets \citep{2006MW,Sutherland,Smullen2016}. Figure \ref{fig:n1_crit} shows the ratio of the planets' semi-major axis to the minimum stable semi-major axis versus the planets period ratio with the binary. Although this pile up of planets might still be explained by observational selection effects \citep{Li2016} alone, the distribution could also arise from unique modes of formation and/or evolution around binaries. In this paper we investigate how mean motion resonances between circumbinary planets can interact with the binary via resonant and secular effects creating additional instabilities in multi-planet resonant systems.

At the current orbit of \Kepler circumbinary planets, dynamical excitation of planetesimals is likely to inhibit planet formation \citep{2007Scholl,Meschiari2012,Rafikov2013,Marzari2013} suggesting that circumbinary planets formed at larger distances and migrated to their current location. More recently, \cite{2015BromleyKenyon} found that circumbinary gas and planetesimals can settle onto streamlines of most circular orbits, facilitating planet formation and a possible solution to the difficulty of formation. Although this effect may assist formation, distant formation and migration remains more likely than in situ formation due to an unrealistically high disk surface density required for in situ formation, even allowing for concentration in streamlines. 

Migration is a natural outcome of the interaction between young circumbinary planets and a circumbinary disk \citep{2012KleyNelson}. Disk driven migration around binaries is similar to that of migration around a single star, but with a hard inner edge set by the inner cavity at around 2 to 3 times the binary separation \citep{Artymowicz1994}. \citet{2008Pierens} and \citet{2007Pierens} studied the formation and migration of circumbinary planets, finding that migrating planets are likely to experience eccentricity growth during migration due to interaction with resonances with the binary. They find that stability during migration is dependent on the final mass of the planet, with Saturn mass planets being more stable than Jupiter mass planets \citep{Kley2014}. \cite{2013Pierens, Kley2014, Kley2015} replicate the approximate orbital parameters of \Kepler circumbinary planets, but find that the final locations are very sensitive to disk structure. 

The migration of multiple planets adds additional complication to the evolution of circumbinary systems. If any single planet migrated to its current location, it would have encountered the N:1 resonances with the binary, a possible source of eccentricity excitation and instability. In multi-planet systems, planet-planet resonances are also likely to exist since disk migration is thought to be a natural origin for resonant exoplanet systems \citep{Lee2002,Mills2016,Trappist}. The multitude of binary-planet and planet-planet resonances, sets up a varity of opportunities to create resonance overlap, which can be a driving source of chaos \citep{1980Wisdom,1999MH,2006MW,Chirikov1979}. Thus, when resonances are spaced such that a body can feel the influence of two resonances at once, orbital instability can result. Two dimensional hydrodynamical models of two migrating circumbinary planets find that scattering events are not only common but consistent with the orbits of the \Kepler circumbinary planets \citep{Kley2015}. In this work we identify the likely origin of such scattering events.

This work is also motivated by the near resonant chain of circumbinary bodies within our own solar system; Pluto, Charon and their 4 circumbinary moons reside near a 1:3:4:5:6 resonant chain. Although the best estimates of the bodies' masses and orbits suggest that none of the resonances are currently active \citep{Showalter2015}, some of the formation scenarios invoke resonant migration of the moons. Current theories \citep{Canup2005} suggest that Charon formed due to an impact with Pluto and that the outer moons formed from the material leftover from the collision. Although the moons most likely formed from the collision, simulations of the Pluto-Charon forming impact \citep{Canup2011} suggest that the collision is not likely to create enough material at significant distance from the binary for the moons to form in their current locations. This conflicts with the fact that their current orbital properties, such as low eccentricity and low inclination, suggest in situ formation is most probable \citep{Brozovi2015}. \cite{Cheng2014} explored another formation scenario that involved capturing inner disc material that moves out via resonant transport. This material becomes caught in a resonance with Charon and then Charon's orbit expands due to tidal interactions with Pluto, moving the moons into their current positions. They found that while this was possible for 1:5, 1:6, and 1:7 resonances with Charon, it was not possible for 1:3 and 1:4 resonances at the location of Styx and Nix, concluding that placement by resonant transport is extremely unlikely.
Additionally, the low eccentricity and inclination of Pluto's moons is easily disrupted by Charon's migration. The moons current positions also rule out certain eccentric tidal evolution scenarios for Charon \citep{2017Smullen} based on the boundary of the inner most stable orbits. The possible interaction of the moons with the orbital resonances we explore here can place even more stringent constraints on their migration histories.

Prospects for finding more circumbinary planets increased earlier this year when the \textit{TESS} mission launched and began discovering new planets \citep{2018Huang_TESS}. \textit{TESS} should discover a number of circumbinary systems although it is less sensitive to circumbinary planets than single star planets due to the difficulty of obtaining multiple transits with a short observing timespan for most of the sky. Because of this, the transit probability of circumbinary planets compared to planets around single stars for \textit{TESS} is smaller than the transit probability for the \Kepler, \textit{K2}, and \textit{PLATO} missions \citep{Li2016}. New methods of transit data analysis \citep{2016WelshOrosz} may also improve yields; at present circumbinary planets have been found by eye, rather than through automated methods due to the inability of typical detection pipelines to account for the binary host stars.

In this paper we investigate the wide range of instabilities that arise in multi-planet circumbinary systems, We begin by calculating analytically the conditions for mean-motion resonance overlap based on the resonance widths of planet-planet and planet-binary resonances in Section \ref{sec:analytic}. In Section \ref{sec:numeric} we develop numerical methods for migrating circumbinary planets in resonance, accounting for the forced eccentricity driven by the binary. In Section \ref{sec:instab} we explain three possible origins of instabilities due to interactions between the binary and a pair of resonant planets and demonstrate these instabilities numerically. We explore the influence of migration on the instabilities in Section \ref{sec:mig}, and discuss how these instabilities are likely to shape the architecture of circumbinary systems and how observed systems can place limits on migration rates in section \ref{sec:discussion}.

\section{Analytic Methods}

Previously developed analytic theory for calculating the location and range of influence of mean motion resonances has assumed mass hierarchies similar to the resonances found in the solar system. The circumbinary systems we are interested in violate these assumptions in two ways. First, previously developed theory assumes the majority of the mass of the system is dominated by one central body rather than two. Orbits about binaries can have substantially non-Keplerian orbits, with substantial forced eccentricities and precession rates. Second, many explorations of resonant bodies assume that one of the bodies in the resonance is significantly more massive than the other, so that the perturbations on the more massive body due to the less massive one can be ignored. These restrictions apply to Jovian resonances in the asteroid belt and the 3:2 resonance between Pluto and Neptune.  In our study, we are interested in mean motion resonances between planets of comparable mass, meaning that additional resonant angles may exhibit libration.

\subsection{Deviation from Keplerian Motion}
\label{sec:analytic}

To accurately identify resonances about a binary, we must account for the deviation of the planets orbits from the standard Keplerian solutions. The locations of both planet-planet and binary-planet resonances shift because of the additional quadropole term in the potential.

We use the analytic theory for orbits around a circular binary developed by
\cite{LeePeale2006} in order to explore the orbits of satellites in the Pluto-Charon system. They show that circumbinary orbits can be represented by the superposition of the circular motion of a guiding center, the forced oscillations due to the non-axisymmetric components of the potential rotating at the mean motion of the binary, the epicyclic motion, and the vertical motion. They  derive modified mean motion as well as the pericenter and nodal precession rates.
The modified mean motion $n_0$ is given by
\begin{equation}
n_0^2 = \Big[ \frac{1}{R} \frac{\textnormal{d}\Phi_{00}}{\textnormal{d}R} \Big]_{R_0}=\frac{1}{2}\Big\{
\frac{m_2}{m_{12}}b^0_{1/2}(\alpha_2)+\frac{m_1}{m_{12}}b^0_{1/2}(\alpha_1)+\frac{m_1 m_2}{m_{12}^2}\frac{a_{12}}{R_0}[\frac{\textnormal{d}}{\textnormal{d}\alpha}b^0_{1/2}(\alpha_2)+\frac{\textnormal{d}}{\textnormal{d}\alpha}b^0_{1/2}(\alpha_1)]
\Big\} n_K^2
\label{eq:meanmo}
\end{equation}
where R is the distance to the circumbinary body from the barycenter, $\alpha_j=a_j/R$, $a_j$ is the semi-major axis of the $j$th body, $n_K=[G m_{12}/R^3_0]^{1/2}$ is the Keplerian mean motion at $R_0$. The Laplace coefficient is defined by
\begin{equation}
\frac{1}{2}b^{(m)}_{1/2}(\alpha_s)=\frac{1}{2\pi}\int_0^{2\pi}\frac{{\rm e}^{-im\psi}}{\sqrt{1-2\alpha_s\cos\psi+\alpha_s^2}}d\psi
\label{eq:laplace}
\end{equation}
and the $j$ derivative of the Laplace coefficient is given by
\begin{equation}
{\cal B}_{1/2}^{(j,m)}(\alpha_s)=\frac{\alpha_s^j}{j!}\frac{\dir^j}{\dir\alpha_s^j}b^{(m)}_{1/2}(\alpha_s).
\label{bigB}
\end{equation}
and finally the expansion in powers of $a_{bin}/R$ of the antisymmetric component of the modified potential equivalent to two rings of mass $m_1$ and $m_2$ at $a_1$ and $a_2$ respectively is given by
\begin{equation}
\Phi_{00} = - \Bigg[1 +
    {1 \over 4 (1+m_2/m_p1)^2}
      \left(m_2 \over m_1\right) \left(a_{12} \over R\right)^2
 +{9 (1 - m_c/m_p + m_c^2/m_p^2) \over 64 (1+m_c/m_p)^4}
      \left(m_c \over m_p\right) \left(a_{bin} \over R\right)^4 + \cdots
    \Bigg] {G m_{12} \over R}.
    \label{Phi00}
\end{equation}
Using the expansion in Equation \ref{Phi00}, Equation \ref{eq:meanmo} is
\begin{equation}
\label{eq:n0}
n_0^2 = \Big[ \frac{1}{R} \frac{\textnormal{d}\Phi_{00}}{\textnormal{d}R} \Big]_{R_0}= \frac{Gm_1 m_2}{64m_{12}R^7} \Big[45 a_{12}^4  (m_1^2 - m_1 m_2 + m_2^2) + 
 48 a_{12}^2  m_{12}^2 R^2 + 64 \frac{m_{12}^4}{m_1 m_2} R^4+...\Big]
\end{equation}

Similarly, the modified epicyclic frequency, $\kappa_0$, is  
\be 
\kappa_0^2 = \Big[ R \frac{\textnormal{d}n^2}{\textnormal{d}R} +4 n^2 \Big]_{R_0}= \frac{Gm_1 m_2}{64m_{12}R^7} \Big[-135 a_{12}^4  (m_1^2 - m_1 m_2 + m_2^2) - 
 48 a_{12}^2  m_{12}^2 R^2 + 64 \frac{m_{12}^4}{m_1 m_2} R^4+...\Big].
\label{eq:k0}
\ee
so that the pericenter precession rate is 
\begin{equation}
\label{eq:precess}
    \dot \varpi = n_0 - \kappa_0.
\end{equation}

We use Equations \ref{eq:n0} and \ref{eq:precess} to calculate the locations of the resonances about a binary including both the shift in mean motion at a given semi-major axis as well as the precession driven by the binary. In reality, the resonance location can also be shifted by the circumstellar disk \citep{Marzari2018} which we do not consider here. We also use the modified mean motion to calculate the state vectors when initializing orbits in our numerical simulations. This offers an improvement for initializing circumbinary bodies in nearly circular orbits. 

\subsection{Resonance Widths}

In order to determine when overlap may occur, we calculate the resonance widths for planet-planet and binary-planet resonances. These widths indicate the maximum deviation from commensurability for two resonant bodies of a particular eccentricity and mass ratio. Not only do we account for shifts in the locations of the resonances due to the presence of the binary, but also we calculate accurate resonance widths in systems with mass hierarchies different than that of resonances in the solar system. \cite{Mardling} presents two expansions of the three body disturbing function valid for arbitrary mass ratios.
The spherical harmonic expansion is valid for well separated systems and for arbitrary eccentricity, while the literal expansion is appropriate for closely spaced systems with non-crossing orbits. We use the literal expansion for calculating the resonant widths of the planet-planet resonances since the relevant resonances are typically low order, $q$, but high degree, $p$, where the ratio of orbital frequencies, $\sigma$, is 
\be \sigma \equiv \frac{\nu_i}{\nu_o} = \frac{p+q}{p} \label{eq:order_degree} .\ee
The literal expansion allows us to calculate the widths of all the angles associated with a resonance for two massive and eccentric planets. Because we are not limited to a circular massive perturber and a massless perturbed body, we adopt the notation of \cite{Mardling} where the subscripts $i$ and $o$ refer to the inner and outer body. There is no assumption that one body is more massive than another. 

The harmonic angle has the form
\begin{equation}
\phi_{mnn'}= n\lambda_i-n'\lambda_o+(m-n)\varpi_i-(m-n')\varpi_o
\label{eq:angle}
\end{equation}
where $\lambda$ is the mean longitude, $\varpi$ is the longitude of the pericenter, $n'$ and $n$ are the integers corresponding to the mean motion of the outer and inner bodies, and $m$ denotes the different angles for a particular resonance. Unless the  period ratio ${\nu_i}/{\nu_o}$ is close to the integer period ratio $n'/n$, the angle will circulate. If the period ratio is close to $n'/n$ and the orbits are able to exchange energy and angular momentum, the angle can librate around a value, typically $0$ or $\pi$. In this case, the conjunctions between the two bodies occurs at nearly the same place. The resonance width identifies the range of period ratios that are distinct from exact commensurability at which this angle can librate. Libration occurs when 
\begin{equation}
\dot\phi_{mnn'}\simeq n\nu_i-n'\nu_o\simeq 0.
\end{equation}
The order of the resonance is $q= n'-n$ and the degree is $p = n'$. 

\subsubsection{N:1 Resonances}

Due to the large mass ratio of the binary, resonances at a significant distance from the binary can have non-negligible widths. These high order, low degree, N:1 resonances are of higher order than those typically explored in the solar system. We use the spherical harmonic expansion from \cite{Mardling} to calculate the widths of the N:1 resonances with the binary. The width of the N:1 resonances is given by Equation 23 of \cite{Mardling},
\begin{equation}
\Delta\sigma_N
=\frac{6{\cal H}_{22}^{1/2}}{(2\pi)^{1/4}}\left[\left(\frac{m_3}{m_{123}}\right)+N^{2/3}
\left(\frac{m_{12}}{m_{123}}\right)^{2/3}\left(\frac{m_1m_2}{m_{12}^2}\right)\right]^{1/2}
\left(\frac{e_i^{1/2}}{e_o}\right)(1-\frac{13}{24}e_i^2)^{1/2}(1-e_o^2)^{3/8}{N}^{3/4}\,{\rm e}^{-N\xi(e_o)/2}
\label{eq:N1}
\end{equation}
where ${\cal H}_{22}$ is an empirical scaling factor equal to 0.71 and $\xi(e_o)={\rm Cosh}^{-1}(1/e_o)-\sqrt{1-e_o^2}$. Equation \ref{eq:N1} predicts $\Delta\sigma_N=0$ when $e_i=0$ suggesting that N:1 resonances are negligible around circular binaries which is not the case. In reality, any non-massless circumbinary body will induce a small eccentricity on the binary via secular interactions making the resonant widths non-zero. Mean motion resonances for massless test particles around circular binaries have non-zero widths according to our numerical simulations. The widths of these N:1 resonances can also be calculated from surface-of-section plots \citep{1998Malhotra,1989Duncan}. We confirm that the widths of the N:1 resonances from surface-of-section plots is similar to that predicted from Equation \ref{eq:N1} with $e_i \simeq 10^{-4}$.  The incorrect prediction of negligible widths from equation \ref{eq:N1} is very likely due to the omission of fast varying terms from the spherical harmonic expansion. The libration frequency of the resonant angle $\phi_N$ is related to the resonant width by
\begin{equation}
\omega_N= \nu_o\Delta\sigma_N/2.
\label{eq:lib_period}
\end{equation}

\subsubsection{Planet-Planet Resonances}

In order to calculate the widths and libration timescales of the resonances between planets of comparable mass, we use the literal expansion in \cite{Mardling}. For a specific resonance, Equation \ref{eq:angle} shows that there are multiple angles associated with the resonance, denoted by $m$, referred to as subresonances. The number of angles is equal to one larger than the degree of the resonance since $n\le m\le n'$. For the N:1 resonances with the binary, there is typically only one relevant angle because of the large mass differences, but for the planet-planet resonances where the masses are comparable, any one of the subresonances can have significant width. For example two planets in the 5:3 resonance, a 3rd degree 2nd order resonance, can have libration in any of the 3 angles, 
$\phi_{335}=3\lambda_i-5\lambda_o+2\varpi_o$,
$\phi_{435}=3\lambda_i-5\lambda_o+\varpi_i+\varpi_o$,
$\phi_{535}=3\lambda_i-5\lambda_o+2\varpi_i$. The relevant angle depends on the mass ratio and eccentricity of the planets. Typically, the angle that includes the term of the less massive body's pericenter will have the most significant width. In the examples in future sections, we will calculate the widths of the $\phi_{mnm}$ angle, $n'=m$, since in our calculations we specify that the inner planet is less massive, while the eccentricities of the planets are similar.

Equation 104 of \cite{Mardling} gives the resonant width from the literal expansion,
so that the width of the $[n'\!:\!n](m)$ resonance is \label{p25}
\be
\Delta\sigma_{mnn'}=2\sqrt{3}\left(\frac{n'}{n}\right)\left|\left[\alpha\,\left(\frac{m_3}{m_{12}}\right)+
\left(\frac{m_1m_2}{m_{12}^2}\right)\right]
\sum_{j=0}^{j_{\rm max}}{\cal A}_{jm}\,F_{mnn'}^{(j)}\right|^{1/2},
\label{reswidth}
\ee
where $\sigma=n'/n$ and $\alpha$ should be replaced by its value at exact
commensurability. ${\cal A}_{jm}$ and $F_{mnn'}^{(j)}$ are the resonance width's dependence on mass ratio and semi-major axis respectively.

\be
{\cal A}_{jm}(\alpha;\beta_2)
=\zeta_m\,\left[\beta_1^{-1}{\cal B}_{1/2}^{(j,m)}(\alpha_1)-\beta_2^{-1}{\cal B}_{1/2}^{(j,m)}(\alpha_2)\right]
\label{Alm}
\ee
where $-\beta_2=m_2/m_{12}=1-\beta_1$,
$\alpha=a_1/a_2$ and $\alpha_s=\beta_s\alpha$, $s=1,2$.
While the dependence on eccentricity is given by
\be 
F_{mnn'}^{(j)}(e_i,e_o)=\sum_{k=0}^j 
(-1)^{j-k}{j\choose k}
X_{n}^{k,m}(e_i)\,X_{n'}^{-(k+1),m}(e_o),
\label{eq:Fmnn}
\ee
where $X_{n}^{l,m}$ and $X_{n'}^{-(l+1),m}$ are the Hansen coefficients defined by
\be
X_{n}^{l,m}(e_i)=\frac{1}{2\pi}\int_0^{2\pi}(r/a_i)^l\,{\rm e}^{imf_i}
{\rm e}^{-inM_i}dM_i= {\cal O}(e_i^{|m-n|})
\label{s4}
\ee
and
\be
X_{n'}^{-(l+1),m}(e_o)=\frac{1}{2\pi}\int_0^{2\pi}\frac{{\rm e}^{-imf_o}}{(R/a_o)^{l+1}}{\rm e}^{in'M_o}dM_o
= {\cal O}(e_o^{|m-n'|}).
\label{s5}
\ee
When calculating $F_{mnn'}^{(j)}(e_i,e_o)$ in Equation \ref{eq:Fmnn}, it is required that $j_{\rm max} >|m-n|+|m-n'|.$ 
This width does not take into consideration the apsidal advance that can contribute to the width of the resonances at very low eccentricities.

\subsection{Secular Effects}

In addition to resonant effects, the binary also induces secular oscillations in planets at a wide range of distances. The spherical harmonic expansion of the secular disturbing function is used to calculate the rates of change of the planets' eccentricities and longitudes of the pericenter. \cite{Mardling} eliminates the terms that vary quickly from the disturbing function in the spherical harmonic expansion to find how the eccentricities and longitudes of the periastra change in Equations 48-51. For the purpose of our study, we assume the eccentricity and pericenter of the binary itself is unchanged by the circumbinary body due to the large mass difference. This simplifies the number of equations we have to integrate to calculate the precession and eccentricity variations to just two equations, 

\be
\frac{\dir e_o}{\dir t}=\nu_o\frac{15}{16}\left(\frac{m_1m_2}{m_{12}^2}\right)
\left(\frac{m_1-m_2}{m_{12}}\right)\left(\frac{a_i}{a_o}\right)^3
\frac{e_i(1+\frac{3}{4}e_i^2)}{\left(1-e_o^2\right)^2}
\sin(\varpi_i-\varpi_o),
\label{eq:dedt}
\ee
\be
\frac{\dir \varpi_o}{\dir t}=\nu_o\left(\frac{m_1m_2}{m_{12}^2}\right)
\left[
\frac{3}{4}\left(\frac{a_i}{a_o}\right)^2
\frac{\left(1+\frac{3}{2}e_i^2\right)}{\left(1-e_o^2\right)^2}
-\frac{15}{16}\left(\frac{m_1-m_2}{m_{12}}\right)\left(\frac{a_i}{a_o}\right)^3
(1+\frac{3}{4}e_i^2)\frac{e_i}{e_o}\frac{\left(1+4e_o^2\right)}{\left(1-e_o^2\right)^3}
\cos(\varpi_i-\varpi_o)\right].
\label{eq:wo}
\ee

For circular binaries, Equation \ref{eq:wo} is identical to  Equation \ref{eq:precess}. We use Equation \ref{eq:wo} to determine the scaling of the pericenter precession from the circular case in \cite{LeePeale2006}. We analytically approximate the pericenter precession as a function of binary eccentricity to find $\dir \varpi/ \dir t$ as a function of $n_0$, $\kappa_0$, and $e_i$,
\be
\frac{\dir \varpi}{\dir t} \approx (n_0-\kappa_0)(1+\frac{3}{2}e_i^2) . \label{eq:wo_us}\ee
We use Equation \ref{eq:wo_us} instead of integrating Equations \ref{eq:dedt} and \ref{eq:wo} for efficiency. 
We also calculate the amplitude of the eccentricity excitation due to secular effects by fitting the results of integrating Equations \ref{eq:dedt} and \ref{eq:wo} and comparing the integration of the equations to N-body simulations. We find

\be \Delta e_o \approx  e_i\frac{a_i}{a_o} \frac{m_1-m_2}{0.4115 \, m_{12}} \label{eq:sec_e} \ee
accurate to 1\% for systems where $e_i < 0.8$ and the sum of the eccentricity excitation and the outer body's initial eccentricity is less than 0.9, $e_{o0} + \Delta e_o <0.9$.   

\section{Numerical Methods}
\label{sec:numeric}

In order to test instabilities in multi-planet circumbinary systems, we use an N-body code to track the evolution of the orbit of the planets as they migrate into resonance and closer to the binary where the instabilities occur. We initialize the planets' starting velocities according to Equation \ref{eq:n0} to more accurately reflect the orbits and eccentricities. We use the N-body code REBOUND \citep{Rebound} to numerically integrate our multi-planet circumbinary systems. We use the IAS15 adaptive, high-order integrator, \citep{IAS15} due to its high accuracy for conservative and non-conservative forces. In order to better describe the non-Keplerian orbits over the course of the system evolution we develop empirical ``modified" orbital elements, see appendix \ref{sec:appen}. These modified or ``geometric" orbital elements are crucial to accurately identifying active resonance, because they better capture the evolution of the argument of pericenter for planets whose orbits are strongly perturbed by the presence of the binary. Computing these orbital elements requires high frequency outputs at the rate of about 100 per orbit, which can slow down the runs and create massive file sizes for long runs. To mitigate this expense, we periodically output orbital data at high frequency at regular intervals. 

When two bodies migrate in resonance, their eccentricities will grow in proportion to the rate of migration \citep{1995Malhotra}. In order to maintain a mean motion resonance while two bodies are migrating at different rates, the bodies need to exchange angular momentum. A byproduct of this exchange is the rapid growth of eccentricity of the bodies describe by Equation 5 of \cite{1995Malhotra},
\begin{equation}
    e_{final}^2 = e_{initial}^2 + \frac{1}{j+1}\ln\frac{a_{final}}{a_{initial}}
\end{equation}
where j is the degree of the resonance. The above equation is only valid in cases where one body is much more massive than the other and the eccentricity growth occurs only in the smaller body, such as Pluto migrating in the 3:2 resonance with a much more massive Neptune. For bodies with comparable mass, the eccentricity of both bodies increases. Migrating resonant planets can easily become unstable due to unchecked eccentricity growth, so some form of eccentricity damping must occur for resonant planets to migrate over extended distances. Migration in a gaseous protoplanetary disk tends to damp eccentricity, increasing stability \citep{2012KleyNelson}. While the magnitude of eccentricity damping will depend on a host of disk parameters, we assume that eccentricity damping is proportional to the rate of migration following \citet{Lee2002,1980GoldreichTremaine},
\begin{equation}
    \dot{e}/e = -K|\dot{a}/a|
    \label{eq:edot}
\end{equation}
where $K$ is a positive constant. A consequence of the proportionality is that the eccentricity reaches an equilibrium value after an initial growth phase, causing the planets to migrate at constant eccentricity for most of their evolution. The equilibrium eccentricity is dependent on the migration rate, $K$, and the total mass of the planets. 

Additional forces in N-body codes can be used to mimic planet migration and eccentricity damping rather than using full hydrodynamic simulations. We modify the migration forces in the planetary migration example in REBOUND, a reproduction of \cite{Lee2002}, so that the velocity is damped to the modified mean motion. Typical forced migration algorithms, which damp to a specified Keplerian eccentricity, are insufficient for modeling circumbinary migration. These methods fail to account for the deviations from Keplerian orbital elements that arise due to the binary. Our migration model works instead by damping to a maximum velocity, characteristic of a specified eccentricity about a binary. The velocity at pericenter is given by \begin{equation}
v_p \approx v_c \, \sqrt[]{\frac{1+e}{1-e}} \label{eq:vp}
\end{equation} where $v_c$ is the circular velocity given by the modified mean motion in Equation \ref{eq:n0}. 
The modified code 
reduces the velocity of the circumbinary body if the velocity is greater than $\sqrt[]{(1+e)}$ times what the circular velocity would be if the current location was the semi-major axis. Because we do not calculate the modified orbital elements during the integration, we do not know the current mean or true anomaly of the body. For a Keplerian orbit, one can trivially invert position and velocity to get semi-major axis and eccentricity, but this is not true for circumbinary orbits. Without phase information (see eqs. 25 and 26 of \citet{LeePeale2006}), one cannot uniquely distinguish between a circumbinary orbit near pericenter as compared to a circular orbit with a higher mean motion and smaller semi-major axis.
We can, however, calculate the maximum velocity as a function of position for given eccentricity by assuming the body is at pericenter. This velocity scaling is similar to the Keplerian case:
\begin{equation}
v_p \approx n_0(a) \thinspace a \thinspace \sqrt[]{\frac{1+e}{1-e}}
\label{eq:vcomp}
\end{equation}
where $n_{0}$ is the modified mean motion at $a=r/(1-e)$, using the modified motion.
Thus to damp to a specific $e$, we measure $v$, calculate $v_p$, set $a = r/(1-e)$, and then only damp the velocity when $v>v_{p}$. 
Using this formalism prevents over damping of the eccentricity, which occurs when one neglects the change in mean motion due to the binary. Note that this only damps the eccentricity when the orbit is actually near its pericenter, and thus the coefficient $K$ in Eq. \ref{eq:edot} slightly overestimates the damping timescale, and slightly enhances the effective migration rate.

The purpose of our damping via additional forces is to be able to control the eccentricity at which the planets encounter the N:1 resonances or secular forcing with the binary. This allows us to calculate the resonant widths at the time of encounter and check for overlapping resonances. We acknowledge that there are many possible sources of eccentricity pumping and damping that can affect a migrating circumbinary planet within a protoplanetary disk. We use this damping technique to test our instability mechanisms, not to indicate the specific locations at which these instabilities will always occur.

\section{Origin of Multi-planet Instabilities}
\label{sec:instab}

Chaos and instability in planetary systems typically arises due to overlapping resonances \citep{1980Wisdom,1999MH,2006MW}. Chaotic behavior occurs when a body is under the influence of multiple resonances simultaneously because the widths of the resonances are similar to their separation. During resonant overlap, the bodies experience a random walk in eccentricity, leading to orbit crossings, close interactions with other bodies, scattering, and sometimes ejection. The resonance widths calculated above can be used to determine if a body is in overlapping resonance regions for a particular semi-major axis and eccentricity. Instability due to resonance overlap occurs in this scenario when two resonances involving the same three bodies overlap, for example both the 4:3 and 5:4 mean motion resonances for two planets around the binary.

We identify three possible routes to instability caused by overlapping mean motion resonances in multi-planet circumbinary systems. In all of the following examples, two planets in resonance with each other become unstable due to secular or resonant interactions with the binary. Both resonances between planets and resonances between planets and the binary can overlap to cause chaos; the eccentricities at locations at which resonance overlap occurs depends strongly on the architecture of the system. We explore the cause of the instability in each route and describe in which systems the instabilities are likely to occur. We also demonstrate each instability numerically, illustrating how resonant circumbinary planets behave while approaching thedesse instabilities.

Although our calculation of overlapping resonant widths using the expansions from \cite{Mardling} is a less rigorous approach than other attempts to explain chaos via overlapping first order mean motion resonances, \citep{Deck2013} we demonstrate numerically, in following sections, that the resonance widths are well described by this expansion and are accurate enough to demonstrate the method of these instabilities even if they do not describe the exact criteria for resonant overlap. Additionally, previous work has only focused on overlapping first order resonances instead of the high order resonances that dominate the systems of interest, leaving the method from \cite{Mardling} as the only simple way to calculate these resonant widths.

\subsection{Planet-Planet Resonance Overlap}
\label{sec:pp}

\begin{figure}
\centering
\includegraphics[scale=.5]{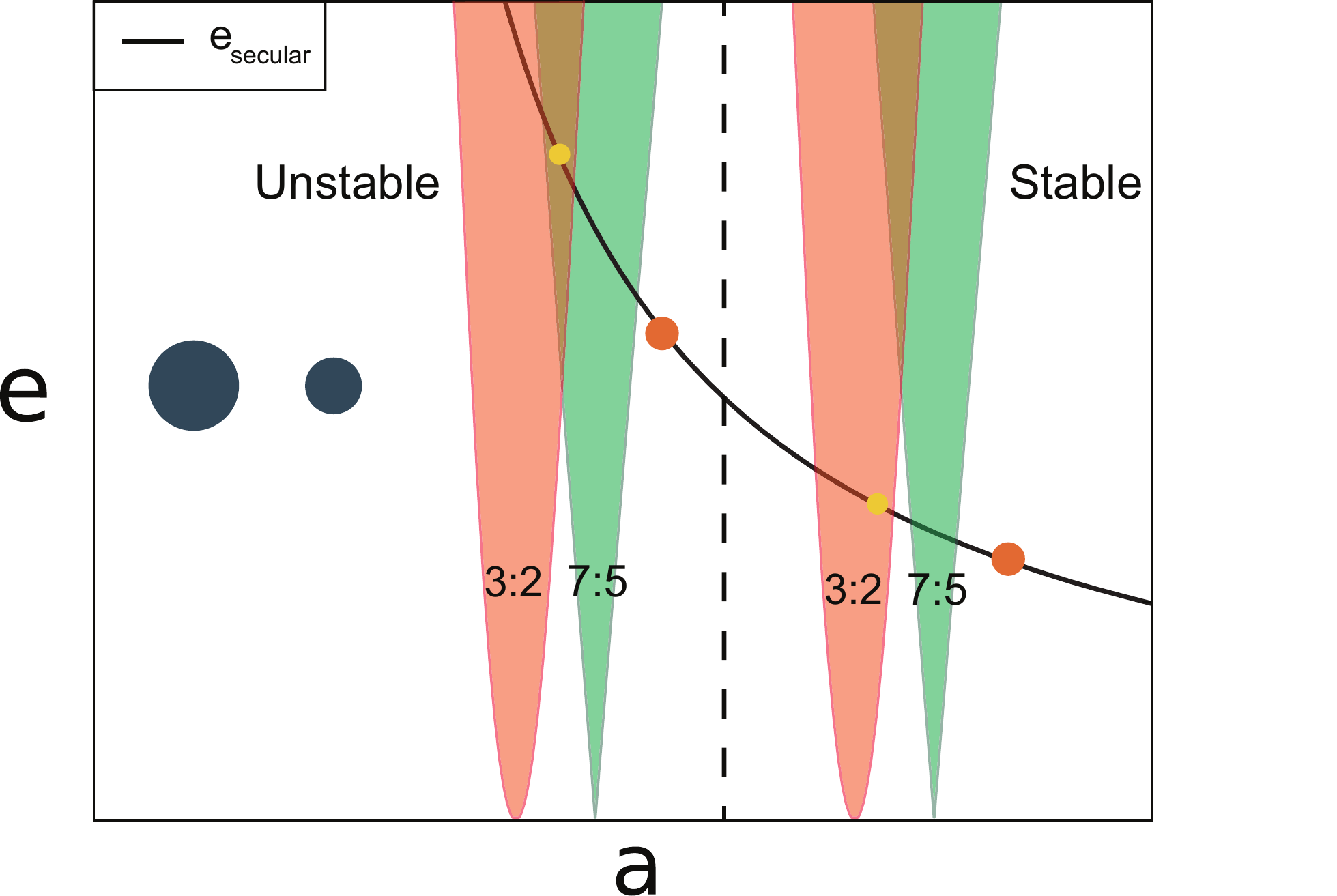} 
\caption{  
Schematic view of planet-planet resonance overlap driven by secular eccentricity variations from the binary. Two pairs of resonant circumbinary planets are shown in the 3:2 resonance. The widths of the 3:2 and 7:5 resonances are shown in red and green. The solid black line shows the maximum amplitude of the secular eccentricity variations driven by the host binary. The dashed vertical line shows the semi-major axis at which the inner planet undergoes secular eccentricity variations large enough to excite the planet into the region of overlap. For the planet pair interior to this critical semi-major axis, yellow inner planet's secular eccentricity variations are significant enough to enter the overlapping region and become ejected while the outer pair has smaller eccentricity variations and the yellow planet remains exterior to the overlapping region and is thus stable.\vspace{.0cm}}
\label{fig:ppcartoon}
\end{figure}

Secular eccentricity variations forced by the binary can drive planets into regions of overlapping planet-planet resonances, PPRs hereafter. Even at significant distances from the binary, the secular variations in eccentricity and argument of pericenter can be large, see Equation \ref{eq:sec_e}. The eccentricity induced by the binary, which increases with decreasing semi-major axis, can ultimately lead to overlap between a two planet-planet resonances as shown in Figure \ref{fig:ppcartoon}. For a circumbinary planet in the 3:2 resonance with a massive outer planet, there exists a critical semi-major axis, interior to which the binary drives up the inner planet's eccentricity until it enters the region that overlaps with the 7:5 resonance. This stability limit depends on the width of the resonances, which are dependent on the mass, eccentricity, order, and degrees of resonance and also the magnitude of the secular effects, which depend on binary mass ratio and eccentricity. Planets will become unstable if they occupy the parameter space in period ratio and eccentricity at which two PPRs overlap.

We calculate the individual resonance widths using Equation \ref{reswidth}, where $m_1$ is the combined mass of the binary stars and the semi-major axis is the distance to the binary barycenter. Although the binary is treated as a single mass in the width calculation, the central position of the resonances are shifted according to the modified mean motion and precession driven by the binary in Equations \ref{eq:n0} and \ref{eq:wo}. The difference in mean motion shifts the spacing between the center of the resonances slightly from the Keplerian values because the mean motion decreases slightly faster than $a^{3/2}$ as in the purely Keplerian case. Equation \ref{reswidth} predicts that resonance widths increase with increasing eccentricity and the separation in terms of semi-major axis decreases with increasing resonance degree, $p$. For example, the 2:1 resonance overlaps with the 3:2 resonance at a larger eccentricity than the 4:3 and 3:2 resonances overlap. Higher order, $q$, resonances have smaller widths, scaling as $e^q$. In general for the planet-planet resonances we investigate, only the first and second order will have significant widths at the relevant eccentricities, meaning that overlap between first or second order resonances will be the cause of instability.

\subsubsection{Numerical Demonstration}
\label{sec:pp_num}

In order to validate the predicted instability, we slowly migrate a pair of planets into resonance using the numerical method described in Section \ref{sec:numeric}.
 The stars have masses $1M_{\odot}$ and $0.4M_{\odot}$, a semi-major axis of 0.1 AU and an eccentricity of 0.2. These values are characteristic of the Kepler circumbinary systems. We initialize two planets with masses $1M_J$ and $0.3M_{J}$, semi-major axes of 3.0 and 4.5 AU on initially near circular orbits, co-planar with each other and the orbit of the binary.  The planets are initially slightly outside the 5:3 resonance. We migrate the outer, more massive planet inward until it enters the 5:3 resonance. The inner planet's eccentricity grows up to the damped value, and then the planets migrate together in resonance at near constant eccentricity. During this time their orbital parameters undergo secular oscillations driven by the binary that increase in amplitude as the planets' semi-major axes decrease. At some distance the secular perturbations reach a large enough amplitude to force the inner planet's eccentricity to a value above the point the 7:5 PPR overlaps with the 9:5 PPR. The planets scatter out of the resonance but migration moves the planets back into the resonance where they scatter again. This process continues until the planets scatter into a resonance of low degree (in this case the 2:1) that is more well separated.

\begin{figure}
\centering
  \includegraphics[width=.4\linewidth]{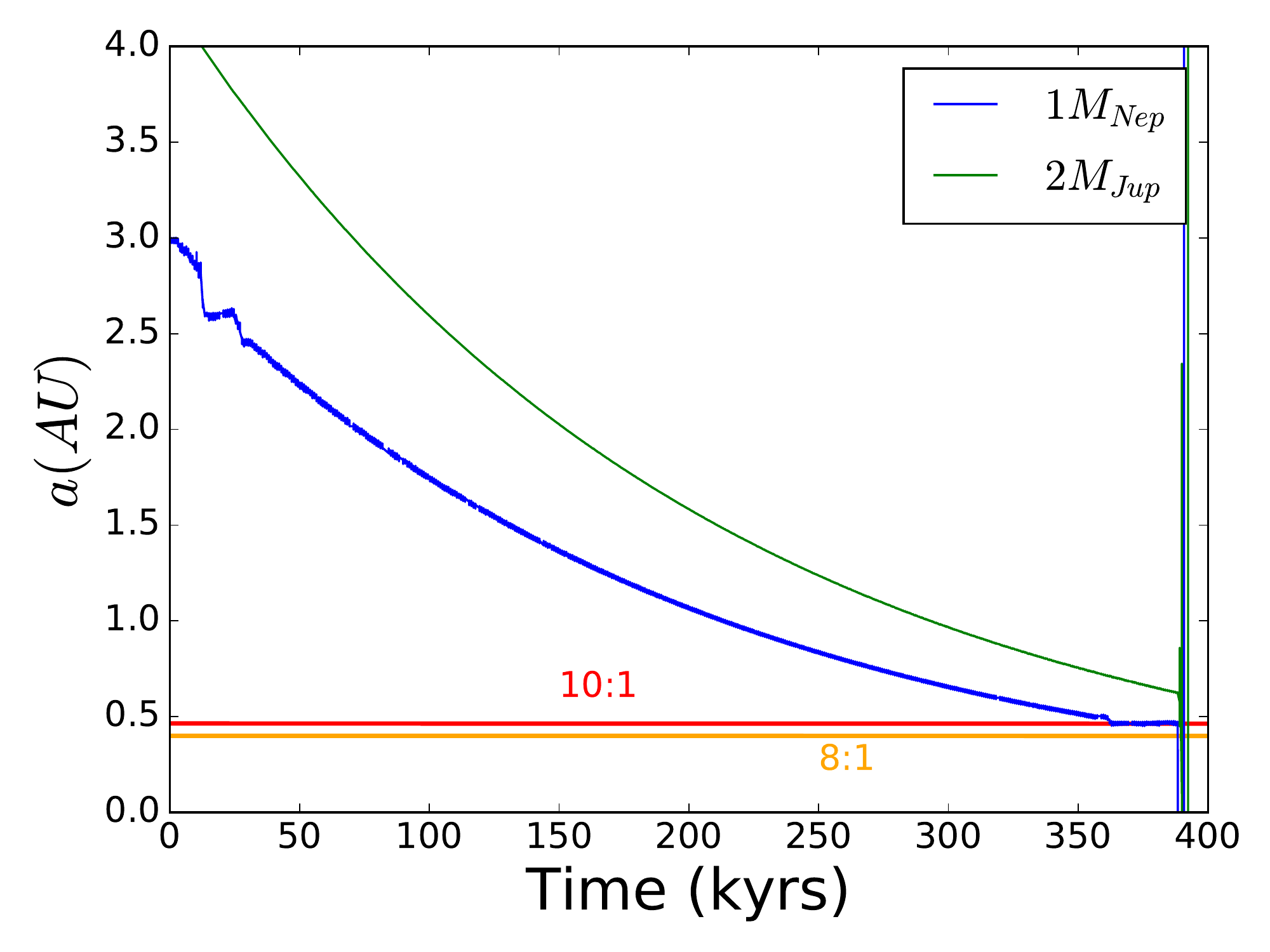}
  \includegraphics[width=.4\linewidth]{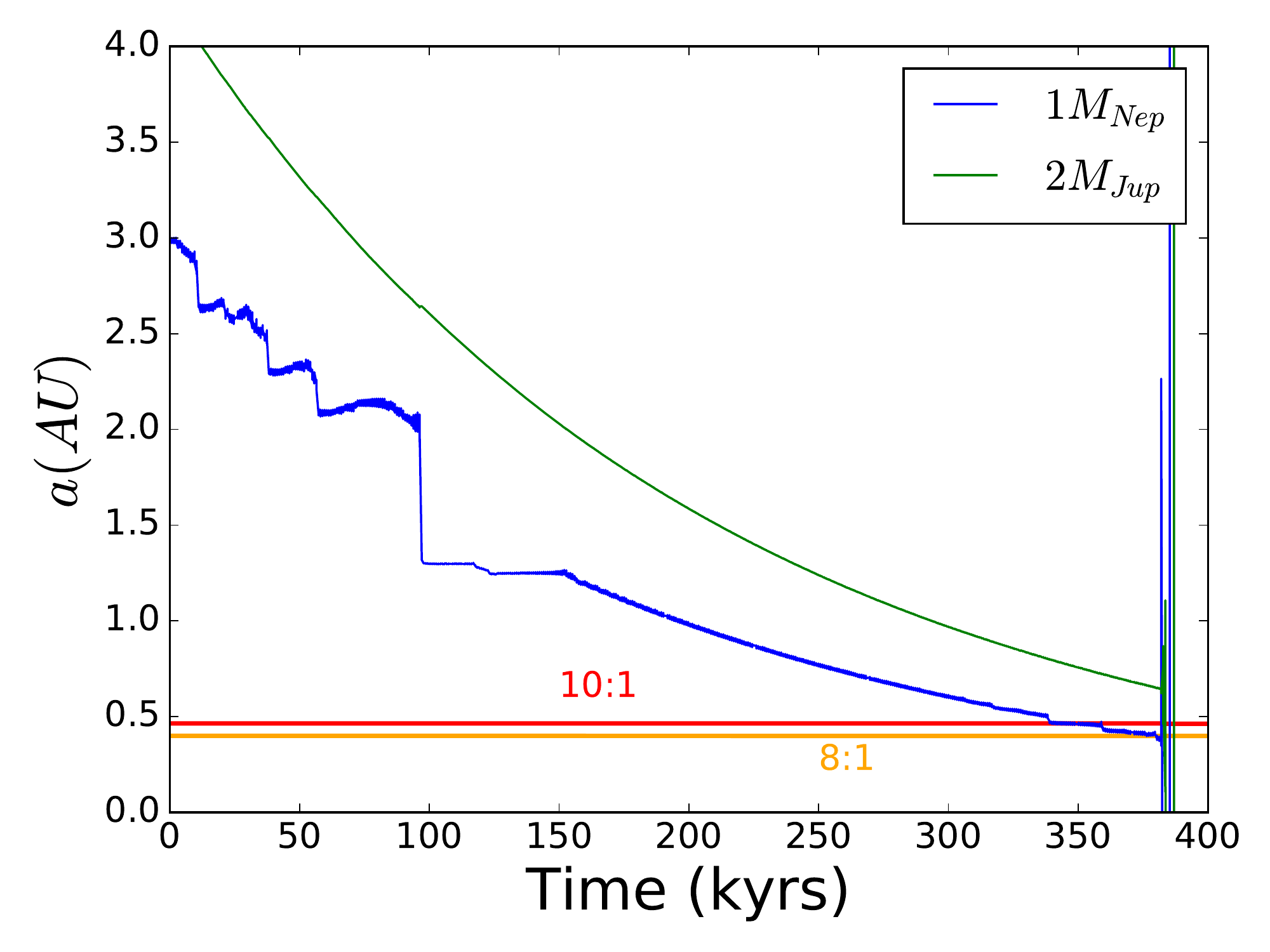} \\
    \includegraphics[width=.4\linewidth]{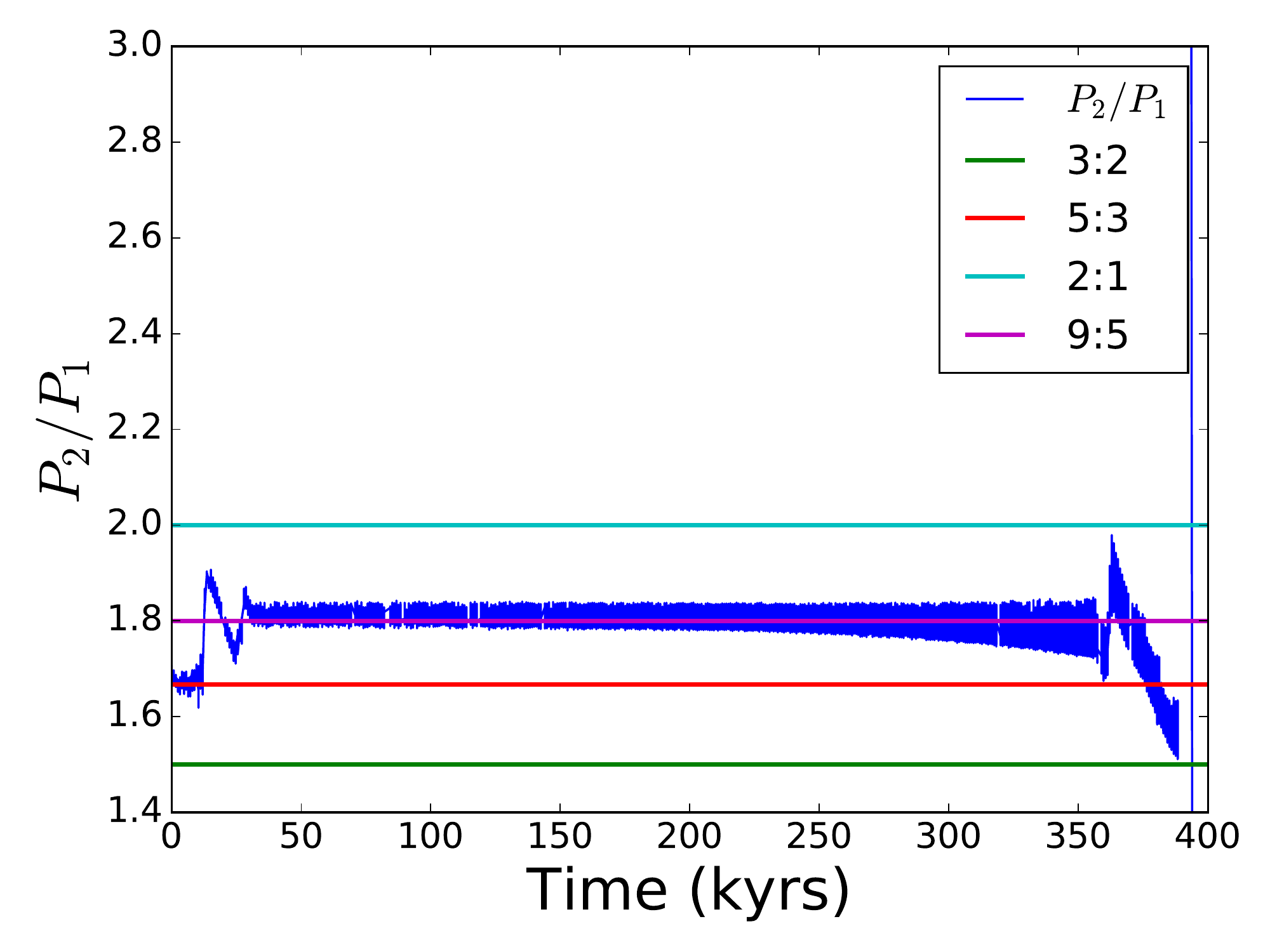}
  \includegraphics[width=.4\linewidth]{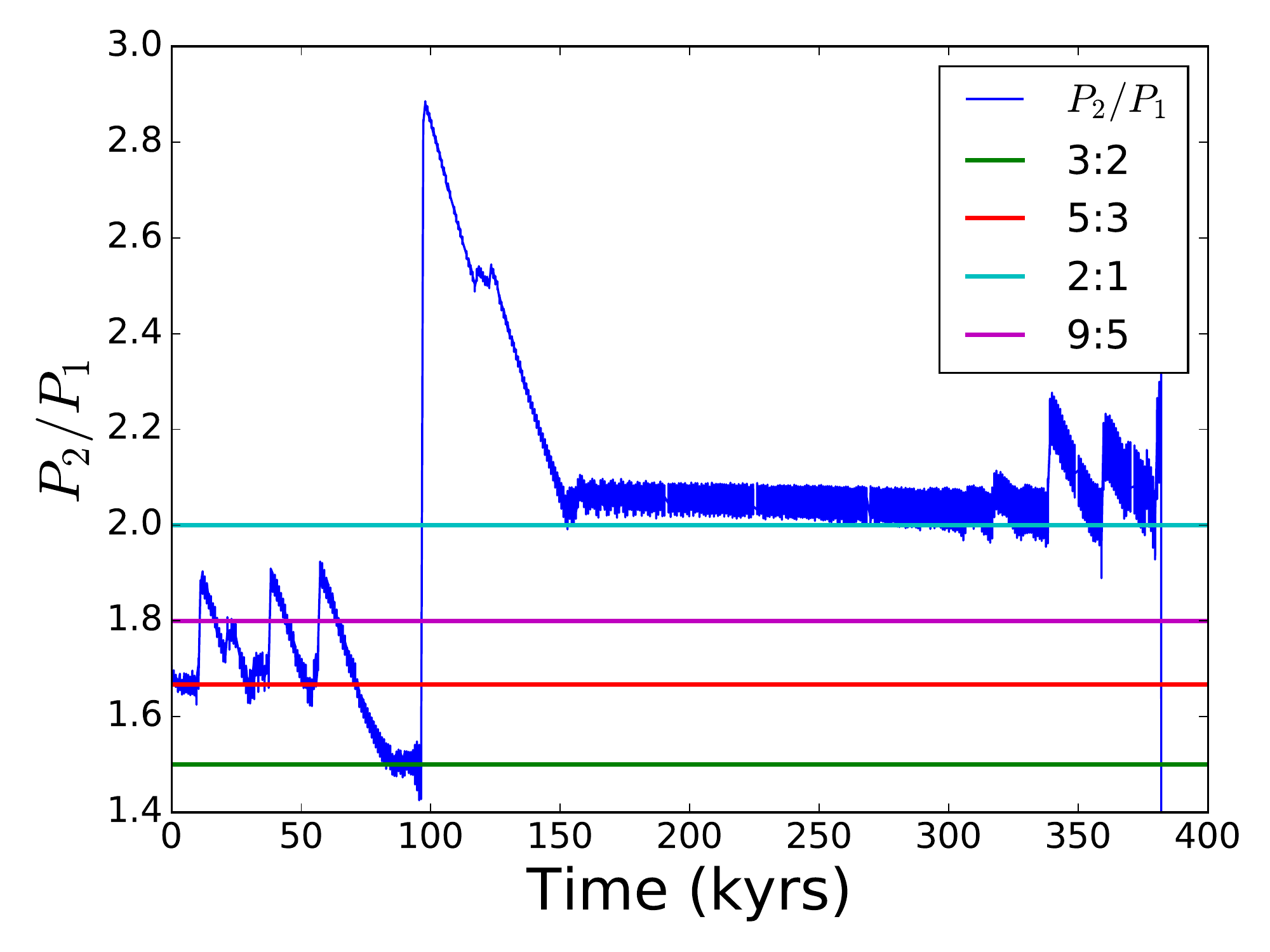}
\caption{The top panels show the evolution of the semi-major axis of a pair of migrating planets, inner planet of 1 Neptune mass and outer planet of 2 Jupiter masses. The location of the 10:1 and 8:1 resonances with the binary are also shown. The bottom panels show the period ratio between the two planets. Two similar runs of planets starting near a 5:3 mean motion resonance. The left shows the evolution around an equal mass binary and the right shows the evolution around an unequal mass binary, both binaries have equal eccentricity. The planet pair around the eccentric binary scatters repetitively when encountering the 5:3 and 3:2 PPRs until encountering the widely separated 2:1 resonance. Once the planets are in a well separated resonance, they are able to migrate to the binary where they encounter the N:1 resonances with the binary where they are ejected as described in the following section.}
\label{fig:pp_num}
\end{figure}

Figure \ref{fig:pp_num} shows the period ratio over time for the planets as they migrate. The right column reflects the evolution described above, while the left column features the evolution where the initial orbital parameters and migration routine are the same except for a change in the mass ratio of the binary. The left shows a binary with an equal mass ratio. According to Equation \ref{eq:sec_e}, the amplitude of the eccentricity variations is zero for an equal mass binary. In this case the planets migrate smoothly in the 9:5 resonance, until a secondary instability (see below) takes over at the 8:1 resonance with the binary. The location at which the PPRs overlap is dependent on the mass and eccentricity of the planets. Lower combined mass of the planets will result in stability at closer proximity to the binary for similar resonances.
The implications of this instability and migration are explored in Section \ref{sec:mig}. 

\subsection{N:1 Binary Resonance Overlap}
\label{sec:N1}

\begin{figure}
\centering
\includegraphics[scale=.5]{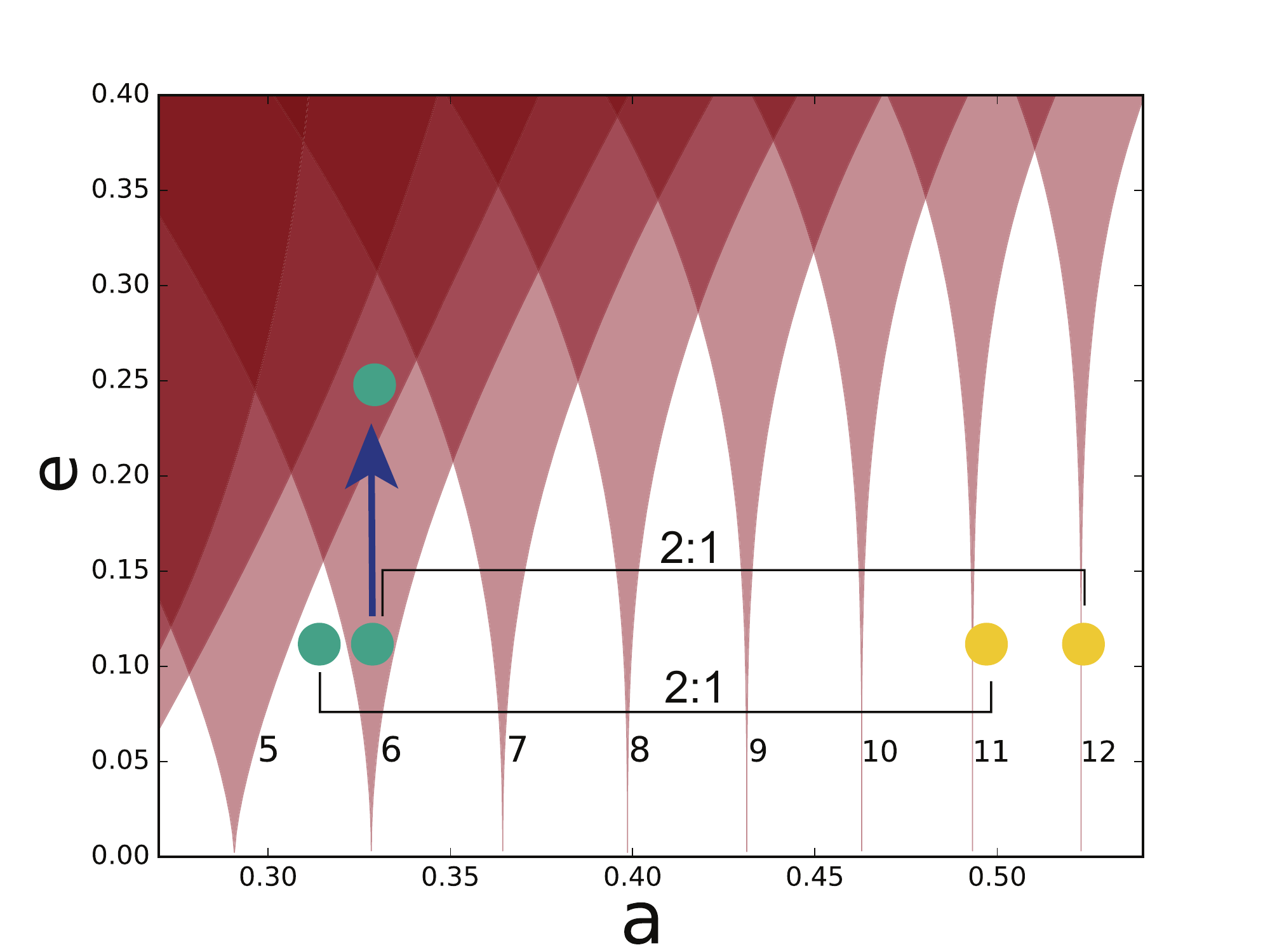} 
\caption{  
The widths of the overlapping N:1 resonances are shown in maroon as a function of planet eccentricity and semi-major axis. The innermost planet of each pair is shown in green while the outer is shown in yellow. The growth in eccentricity due to the interaction with the 6:1 resonance is represented with a blue arrow. The inner planet is scattered once it reaches the point of N:1 resonance overlap while the other planet pair is stable since the inner planet does not overlap with any N:1 resonance and therefore no eccentricity growth occurs.}
\label{fig:n1_cartoon}
\end{figure}

The second instability we explore occurs when migrating resonant planet pairs make one of the planets  susceptible to eccentricity pumping by an N:1 resonance with the binary.
The N:1 resonance can then increase the planet's eccentricity until two N:1 resonances begin to overlap, leading to ejection. For low eccentricity and near equal mass binaries, these resonant perturbations are more powerful than the secular ones. Equation \ref{eq:N1} shows that the widths of the N:1 resonances quickly increase with deceasing N. Important N:1 resonances for typical binaries range from the 5:1 to the 15:1, mostly dependent on eccentricity of the binary and planets and only very weakly dependent on the mass of the planets. The width of N:1 resonances can be seen in Figure \ref{fig:n1_cartoon} with the darker shaded regions showing the overlap of these resonances.

While a single planet of the same eccentricity can survive in an N:1 resonance for a significant time \citep{Sutherland}, the combined resonant forcing due to a PPR and the N:1 leads to eccentricity growth, causing the planet to enter the region of N:1 resonance overlap. This results in a rapid ejection. Figure \ref{fig:n1_cartoon} shows two pairs of planets, both in a 2:1 PPR. The pair with the inner planet in the width of the 6:1 resonance with the binary will have its eccentricity excited from being in both the 6:1 with the binary and the 2:1 PPR until it reaches the point of resonance overlap with the 5:1 binary resonance, which then leads to ejection. The pair with the inner planet slightly interior to the 6:1 will not become disrupted. 

Unlike the instability described in Section \ref{sec:pp}, there is not a critical semi-major axis beyond which this instability occurs. The instability is prevalent for a wider range of binary and planetary parameters as the inner planet's semi-major axis decreases because the widths of the resonances increase, making it easier to overlap at lower eccentricity. This instability is also more likely to occur in binary systems with weaker secular eccentricity oscillations and well separated PPRs, since PPRs overlap due to secular effects, Section \ref{sec:pp}, can occur at any point, not only in the N:1 resonances.

While other resonances exist between the N:1 resonances, such as the 2N$\pm$1:2, these higher order, higher degree resonances are not as strong. Typically the width of the resonance goes as $e^q$, so the 11:2 which is located between the 5:1 and 6:1 is a 9th order resonance and therefore much weaker than the 4th and 5th order 5:1 and 6:1 resonances.

\subsubsection{Numerical Demonstration}
\label{sec:n1_numerical}

We migrate a pair of planets in resonance from a large distance from the binary through the N:1 resonances. The initial set up is similar to \ref{sec:pp_num} with the following changes; the binary is near equal mass so that secular eccentricity variations are minimized and planets are initially just outside of a 2:1 resonance so PPRs are more well separated. The innermost planet's eccentricity is damped to and librates around $e = 0.2$ until encountering the 7:1 resonance. The 7:1 resonance grows the eccentricity of the planet but is not able to grow into the overlapping resonances, thus the planet is able to pass through the resonance without becoming unstable as the eccentricity does not reach the point at which the 7:1 and 6:1 resonances overlap. The planet's eccentricity returns to 0.2 after exiting the resonance. It then encounters the 6:1 resonance and the eccentricity grows until it reaches the point of overlap with the 5:1 and is ejected.

\begin{figure}
\centering
\includegraphics[scale=.4]{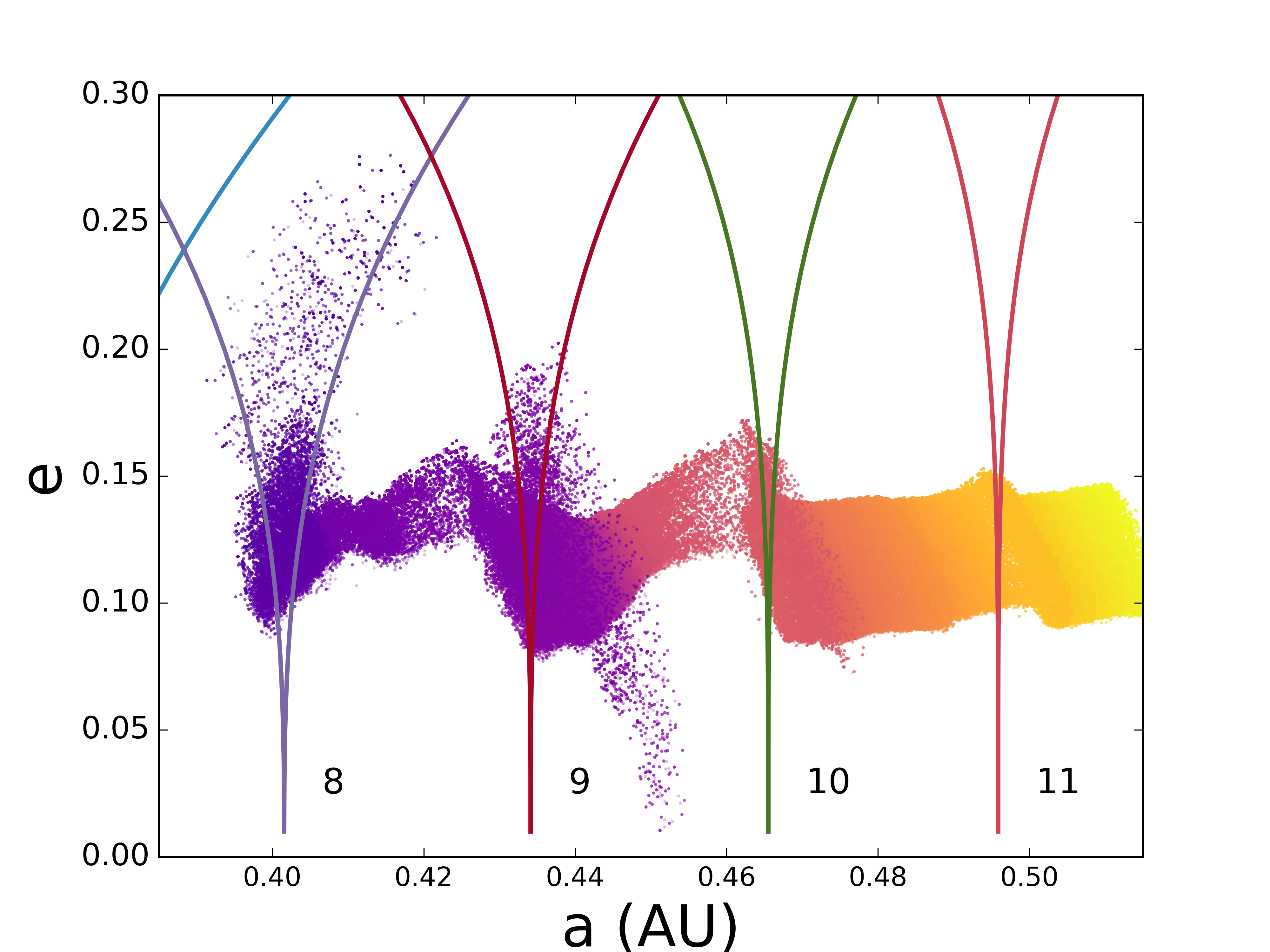} 
\includegraphics[scale=.5]{5b_new_c_3}
\caption{  
\textbf{Left:} The evolution of a planet migrating inward in eccentricity-semi-major axis space while in resonance with an outer planet, colored by time. The planet migrates while damped to e=0.16 until it encounters the 8:1 resonance with the binary where it is ejected. \textbf{Right:} The semi-major axis and eccentricity of a migrating planet as it encounters the N:1 resonances. The N:1 resonances excite eccentricity growth of the planet, shown in red. The planet is able to migrate through a number of N:1 resonances before the eccentricity grows to the point of N:1 overlap. After encounters with the 11:1-9:1, the eccentricity is damped back to 0.16. When it encounters the 8:1 resonance overlap occurs and the planet  is ejected. This instability is largely independent of planet mass.}
\label{fig:n1_num}
\end{figure}

\subsection{Planet-Planet Resonance Overlap due to Eccentricity Pumping from Resonances with Binary}
\label{sec:combo}

\begin{figure}
\centering
\includegraphics[scale=.6]{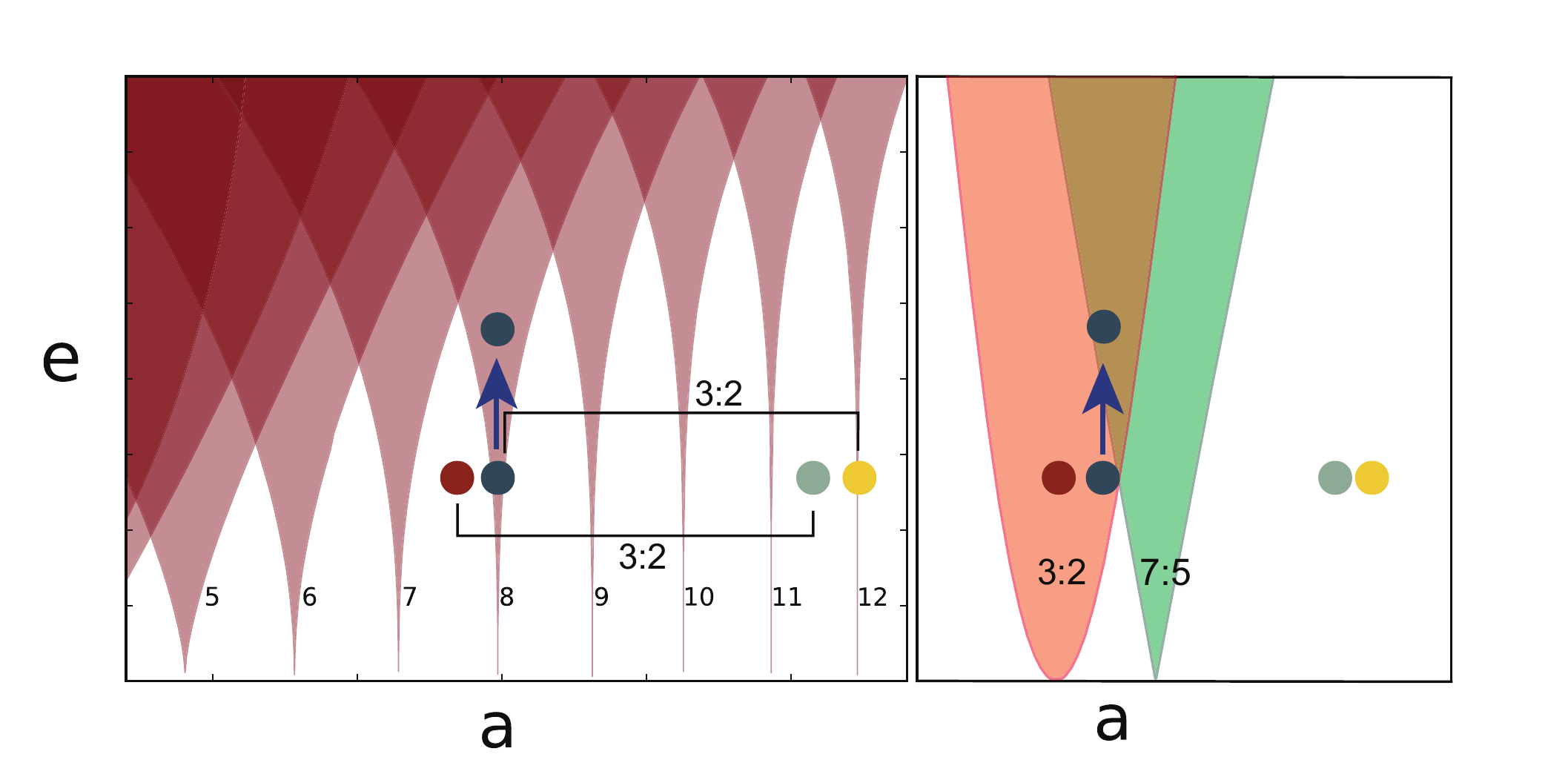} 
\caption{  
Right: N:1 resonance widths, similar to Figure \ref{fig:n1_cartoon}. Left: PPR for two circumbinary planet pairs, similar to Figure \ref{fig:ppcartoon}. Both figures show two planet pairs in a 3:2 PPR. The dark blue and yellow planet pair is unstable since the inner planet overlaps with the 8:1 resonance. This pumps the eccentricity of the planet to the point of 3:2 overlap with the 7:5 PPR. The red and gray planet pair is stable because it does not reach the point of PPR overlap since it is outside the N:1 resonance.}
\label{fig:combo_cartoon}
\end{figure}

PPR overlap described in Section \ref{sec:pp} happens at large distances from the binary but with closely separated planet-planet resonances while instabilities due to N:1 resonances overlap, as described in Section \ref{sec:N1}, happen close to the binary in more well separated planet-planet resonances with weak secular effects. An additional mode of instability may occur when the excitation due to the interaction with N:1 resonances increases the planetary eccentricity sufficiently to cause the planet-planet resonances overlap. 
We expect this mechanism to be relevant when the innermost planet is close to the binary, but the secular effects are weak due to low binary eccentricity or equal mass ratios. As shown in Figure \ref{fig:combo_cartoon}, the inner planet in a 3:2 resonance with an outer planet would become unstable if it is within the width of a N:1 resonance, since the resonant perturbations of the binary will excite the inner planet into the overlapping region of the 3:2 and 7:5 planet-planet resonances.

Unlike the instability described in \ref{sec:N1}, the inner planet does not enter the region of overlapping N:1 resonances, but reaches the critical eccentricity for overlap of the planet-planet resonance.
This instability, like the one described in \ref{sec:N1}, does not have a critical semi-major axis where the instability happens, but instead is limited to the widths of the N:1 resonances. Since the widths of the N:1 resonances increases with decreasing N, the resonances are more relevant for smaller N. The maroon and light gray planet pair in Figure \ref{fig:combo_cartoon} is safe between the 7:1 and 8:1 resonances while the dark blue and yellow planet pair is unstable because of the dark blue planet's location within the width of the 8:1 resonance. 

Whether the instability in \ref{sec:N1} or the one described here should dominate depends on whether the N:1 resonances or PPRs overlap at a lower eccentricity. The relevant eccentricity for excitation depends on the mass ratio and eccentricity of the binary as well as the order, degree, and mass ratio of the planets. For a particular system, which N:1 the planet interacts with will determine which route of instability dominates. For slowly migrating systems with higher degree resonances, this route should be more likely than the instability in \ref{sec:N1}, since the N:1 overlap happens at large eccentricity. Because PPR overlap is likely to scatter poorly separated resonances before encountering strong binary resonances, this instability should be relatively rare. Determining exactly which PPRs are responsible for ejection is complicated by the large number of PPRs, of higher degree and order, that have significant widths when the N:1 resonances excite the eccentricity rapidly. We carried out a large number of n-body integrations to attempt to validate this mode numerically, and while systems went unstable in a manner inconsistent with the instabilities in 4.1 and 4.2, we could not identify a unique source of the instability.

\section{Role of Migration}
\label{sec:mig}

\subsection{Disk Migration}

In the previous section, we consider planets migrating slowly at a fixed rate from a large distance away from the binary to locations where the instabilities can occur. Now we consider the physical origins of that migration and determine the typical rates of migration. After a planet forms in a circumstellar disk, there can be time for the disk and planet to interact before the disk dissipates. The consequence of disk-planet interactions is the migration of planets. The differential migration speeds, which are dependent on the mass of the planet and semi-major axis, are one source of PPRs. Circumbinary planets are not thought to have formed in situ but rather formed at larger semi-major axes and migrated to their current locations \citep{2007Scholl, Meschiari2012,Rafikov2013,Silsbee2015}. Planetary migration  not only forms PPRs but also transports planets into the region where secular eccentricity oscillations and N:1 resonances become destructive.

The impact of disk migration on the architecture of the final planetary system is dependent on the planets migration rates, particularly for interactions driven by encounters with N:1 resonances. A planet will be able to safely migrate through a given N:1 resonance if the resonant crossing time, the width of the resonance divided by the migration rate, is less than the libration period \citep{Lithwick2008}. In this case, the change in orbital elements due to the migration is greater than the change induced by the resonance, and thus the planet does not undergo as much eccentricity growth. If the crossing time is similar to the libration time, the planet may become temporarily trapped in the N:1 resonance before passing through. During this time the eccentricity can grow, as in Figure \ref{fig:n1_num}, but the interaction time will still typically not be enough for the N:1 to excite the eccentricity of the planet into the overlapping region. 

The rate of migration in real systems of course depends on the characteristics of the disk and planet, such as viscosity and mass, that are subject to change as the disk evolves and the planet accretes. To demonstrate the importance of the ratio of the resonance crossing time to libration time, we adopt a constant migration rate, $\frac{\dot{a}}{a}$. Although a circumbinary disk will have a different structure and migration rate than a disk around a single star, such as a central cavity near the binary \citep{Artymowicz1994}, we can still use single star migration rates as a reasonable benchmark for real systems. \cite{1980GoldreichTremaine} present the Type I migration rate for a planet around a single star as
\begin{equation}
    \frac{\dot{a}}{a} = -1.67\Big(\frac{M}{M_0}\Big)\Big(\frac{\Sigma a^2}{M_0}\Big)\Big(\frac{a}{\Delta}\Big)^3\Omega
    \label{eq:typei}
\end{equation}
where $M$ is the mass of the planet, $M_0$ is the stellar mass, $\Sigma$ is the gas surface density, $a$ is the semi-major axis, $\Omega$ is the orbital frequency, and $\Delta =  2a/3m_{max}$ where $m_{max}$ is the azimuthal wavenumber. The general scaling for Type II migration where the planet has opened a gap in the disk is the viscous timescale \begin{equation}
    \frac{\dot{a}}{a} \simeq \frac{3\nu}{2a^2}=\frac{3}{2} \alpha \Big(\frac{H}{a}\Big)^2 \Omega = 5.3\times 10^{-5}\Big(\frac{\alpha}{4\times10^{-3}}\Big)\Big(\frac{H/a}{0.05}\Big)^2\Big(\frac{M_0}{0.32 M_{\odot}}\Big)^{1/2}\Big(\frac{a}{AU}\Big)^{-3/2} yr^-1
    \label{eq:mig_rate}
\end{equation}
from \cite{1997Ward} where $\nu$ is the viscosity, $\alpha$ is the dimensionless viscosity parameter, $H$ is the scale height, $a$ is the semi-major axis, $\Omega$ is the orbital frequency, and $M_0$ is the total stellar mass. While real systems will have rates that deviate from these values (see e.g. \citet{Duffell2014}), our goal here is to simply demonstrate the relevant order of magnitude rates expected in circumbinary systems. For more recent formulas more accurately describing the total torque of the disk on the planet, see \cite{Paardekooper2011}.

\begin{figure}
\centering
\includegraphics[width=115mm]{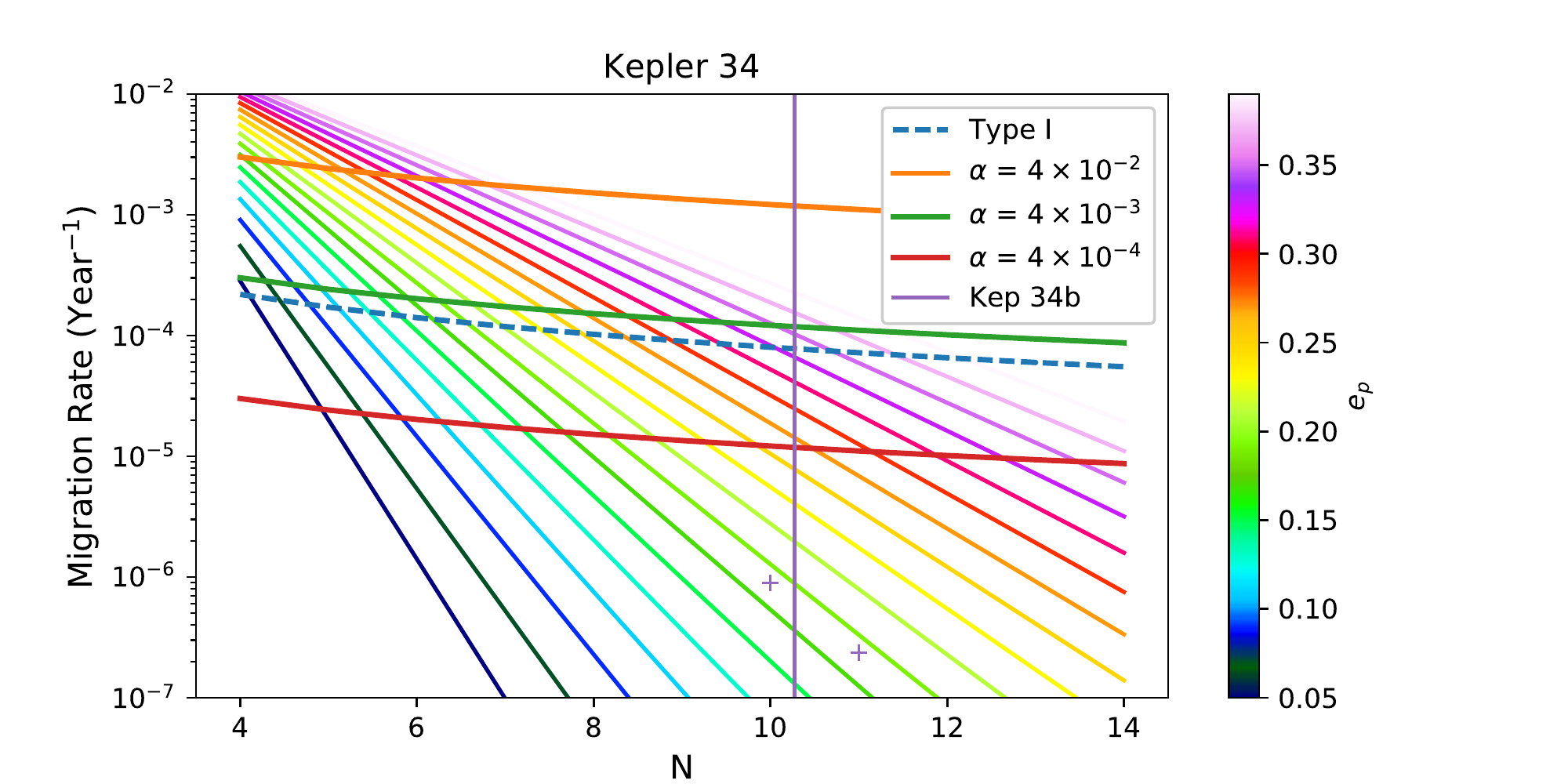}
\caption{  
Critical migration rate at N:1 resonances for a range of planet eccentricities for migrating circumbinary planet pairs around binary stars similar to Kepler 34 with masses of $1.05M_{\odot}$ and $1.02M_{\odot}$ with $0.23$ AU separation and $e_{bin}=0.52$. The range of planet eccentricities are shown as different colors. The area above the diagonal lines is the stable region while the area below the lines is unstable due to slower migration rates. Migration rates for type I \citep{1980GoldreichTremaine} and type II \citep{1997Ward} for a number of viscosities are shown for comparison. The location of Kepler 34b at $1.09$ AU between the 10:1 and 11:1 resonances is shown as a vertical line. Two crosses show the required migration rates for Kepler 34b with an eccentricity of $0.18$ to pass through the 10:1 and 11:1 resonances while in a resonance with an outer planet.}
\label{fig:mig}
\end{figure}

Using the migration rate from Equations \ref{eq:typei} and \ref{eq:mig_rate} along with the resonance width from Equation \ref{eq:N1} we can find the resonance crossing time and then compare this timescale with the libration period, $2\pi/\omega_N$, from Equation \ref{eq:lib_period} to determine at which N:1 the instability described in Section \ref{sec:N1} should occur. Figure \ref{fig:mig} shows the the migration rate needed to safely pass through each N:1 resonance. The multi-color diagonal lines show the critical migration rate for different planet eccentricities. Migration rates from Equation \ref{eq:mig_rate} for typical disks for a number of $\alpha$ values are shown intersecting the lines of different planet eccentricity. The intersection between the two lines show at what N:1 the planet becomes unstable.

Figure \ref{fig:mig} shows that the critical migration rates are close to the predicted rates for Type I and II migration for a range of disk parameters. This indicates that migrating resonant circumbinary planets are likely to encounter this instability. Since typical migration rates are near the critical rates, this instability could be a barrier to planets migrating to their current location as well as a diagnostic tools for measuring migration rates or eccentricity damping of disks for resonant planets observed interior to these N:1 resonances, discussed more in Section \ref{sec:discussion}.

\subsubsection{Numerical Results}
To demonstrate this evolution numerically, we initialize two planets at near zero eccentricity just outside of a 2:1 resonance, both far away from the binary with the inner planet's initial semi-major axis greater than 30 binary separations. The planets migrate into resonance and then continue migrating towards the binary. If the binary has significant eccentricity and an unequal mass ratio, the first instability encountered is the one described in Section \ref{sec:pp}, resulting in the planets scattering out of the PPR. Scattering, rather than ejection is the typical outcome of PPR overlap. The inner, less massive planet can scatter inward and be damped by the disk, only to be captured into resonance again by the outer planet as the outer planet continues to migrate after the scattering event. The planet can be scattered again if the new resonance is not well separated from nearby PPRs. This continues until the planet is either ejected or is scattered into a well separated resonance, such as the 2:1 PPR. 

\begin{figure}
\centering
\includegraphics[width=100mm]{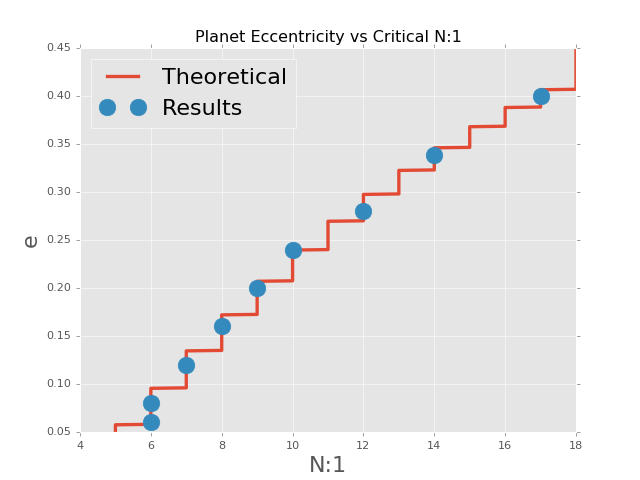}
\caption{ The critical N:1 resonance for a fixed migration rate for a range of planet eccentricities. Predicted unstable N:1 (blue) compared to numerical integration(orange) for a number of eccentricities. Theoretical points outnumber numerical results as we have yet to run a full set of simulations at all ranges of eccentricity.}
\label{fig:n_e_mig}
\end{figure}

In order to test the instability predictions of the previous section, in practice we hold migration rates fixed and instead vary the planet eccentricity damping. Since the widths of the resonances are very sensitive to the eccentricity of the planets, changing the value that the eccentricity is damped to results in a larger range of critical N:1 resonances. We then compare the point at which the inner planet becomes unstable with the prediction from Figure \ref{fig:mig}. The results are shown in Figure \ref{fig:n_e_mig}.

The blue points show the predicted unstable N:1 by comparing the resonance crossing time to the migration rate. Each orange point shows the location the inner planet before becoming unstable due to interacting with an N:1 resonance.  The results from the numerical integrations are near the predicted value or at a slightly greater location. We attribute the modest differences between the numerical results and the theoretical predictions to the somewhat inexact eccentricity damping. 

\subsection{Binary Orbit Evolution}

While we have previously discussed disk migration as the means to drive planets towards the aforementioned instabilities, evolution of the binary orbital parameters can have much the same effect. Circumbinary hosts likely undergo very large scale orbital evolution after forming at wider separations and shrinking their orbits via mechanisms including migration through disks, gravitational interactions with other stellar companions in triple systems, and dissipation due to massive gas reservoirs \citep{Bate1995,Bate2002,Kratter2010a,Offner2010,Kratter_Moe_2017}. Presumably this occurs prior to the formation of the planets \citep{Kratter2017,2015Hamers,2015Munoz}. On longer timescales, the orbits of the close binaries can further evolve through slow tidal dissipation \citep{1981Hut}. Naively, one might expect this to merely stabilize a system because tidal dissipation will tend to decrease the binary eccentricity and semi-major axis, both of which will reduce the secular eccentricity oscillations driven in the planets. However, as the orbit of binary stars evolves, the N:1 resonances are shifted relative to the orbiting planets, potentially moving planets into the N:1 resonances where the above instabilities can occur. 

For example, a resonant pair of planets can quickly migrate in a circumbinary disk through N:1 resonance before stopping between two N:1 resonances near the binary. The termination of migration could be caused by the dissipation of the disk or reaching the disk boundary. Later, close binaries that migrate due to tides might push the planets back through a given N:1 resonance. Since the migration rate of the binary is much slower than the previous planet disk migration, on the order of $10^8$ to $10^9$ years, the system will always be in the regime where the resonance crossing time is very slow compared to the libration time, and thus the planets will always be excited to the maximum eccentricity provided by the N:1 resonances. We defer detailed follow up of this mechanism to Sutherland et al, in prep.

\section{Implications and Discussion}
\label{sec:discussion}

We have described a number of instabilities due to mean motion resonance overlap in multi-planet circumbinary systems, reproduced the instabilities numerically, and demonstrated that critical migration rates for undergoing these instabilities are likely to be near expected disk migration rates. The location of the instabilities depends on a large number of factors including the mass ratio, semi-major axis, eccentricity of both the binary and the planets, the resonance order and degree, and for the case of migrating planets, the migration rate which is in turn dependent on the disk parameters and planet mass. Despite the large number factors, understanding these instabilities can serve as a test of migration rates, likelihood of the discovery of resonant planets, maximum eccentricity growth during migration, and prospect of discovering planets at particular locations.

\subsection{Lack of High Degree Resonances around Eccentric Binaries}
\label{sec:pp_discus}

Because of the large secular perturbations of eccentric binaries at large distances, only planets in well separated resonances should exist close to the binary. The strongest predictor of planet location from the previously mentioned instabilities is due to the overlap between PPRs and the secular effects of the binary (Section \ref{sec:pp}). 
Note that resonance overlap can occur between both first order and second order resonances, increasing the pathways to disrupt resonant planetary systems. 
As the planets migrate towards the separations at which the Kepler systems currently reside, often the only resonance well separated enough to remain stable is the lowest order and degree resonance, the 2:1 PPR. Thus, surveys should not find close-in planets in mean motion resonances other than the 2:1 for sufficiently eccentric binaries. 

\subsection{As a Test of Planet Migration Rates}

We have shown that if a resonant planet pair encounters an N:1 binary resonance for a time longer than the libration period, the planet will be ejected from the system after being pumped to eccentricities of N:1 overlap (Section \ref{sec:N1}). The majority of \Kepler circumbinary planets are interior to the 11:1 resonance, Figure \ref{fig:n1_crit}. If the planets migrated to their current location while in resonance with additional outer planets, we can estimate the minimum migration rate these planets would need to safely migrate through the N:1 resonance to which they are interior. This process is not directly dependent on outer planet mass, eccentricity, or the degree and order of rhe PPR, unlike the case discussed in Section \ref{sec:pp_discus}). This instability only requires the planet to be in some PPR with an outer planet since the PPR resonance is responsible only for modest excitement of the eccentricity, while the N:1 overlap is responsible for the rapid ejection. The critical migration rate is determined by the binary characteristics and the inner planet eccentricity and location.

For each \Kepler circumbinary planet, and for only the innermost planet for the only known multi-planet system, we determine the minimum migration rate that a planet would need if it had migrated to its current location in resonance with an undiscovered outer planet. For some planets that are further away from circular binaries, such as \Kepler 453b, this rate is very low, while for others such as \Kepler 64b and \Kepler 34b the minimum migration rate is near $\frac{\dot{a}}{a}\simeq 10^{-6} yr^{-1}$. While a large caveat is that there is no evidence to suggest these planets must have migrated in resonance, the largest minimum migration rates are within expected rates for type I or II migration suggesting that the \Kepler systems are consistent with migration while in resonance. Additionally, our numerical simulations show that planets do not always immediately get ejected when encountering the critical N:1 resonance. If the planet is initially on a circular orbit and the migration rate is fast, eccentricity growth will occur but not reach a point of overlap before exiting the N:1 resonance. At this point the planet may be damped back down to its initial eccentricity as in Figure \ref{fig:n1_num} or the eccentricity can continue to grow as the planet encounters additional N:1 resonances. Imperfect damping can make planets migrate at a range of eccentricities, affecting the resonance encounter time, the rate of eccentricity growth, and the likelihood of resonance overlap.

In the future, resonant circumbinary planets may be discovered that can be used for a more direct test of migration rates. Resonant planet pairs are not outside of the discovery space for \textit{TESS} and radial velocity surveys. Additional non-transiting planets in resonance with transiting planets can be discovered via transit timing variations. The TTV signal increases dramatically for planets in resonance \citep{2005HM} indicating that TTVs might be a rich source of resonant planets pairs. If the orbits and masses of the binary stars are well defined, the crossing time can be calculated and if the eccentricity of the planet is known, a critical migration rate can be inferred. Although all planets may not have migrated to the maximum distance inward past the N:1s, a large sample of resonant planet pairs near low order N:1s should provide a minimum migration rate. 

It is possible that the eccentricity of the planets during migration was larger than the current discovered value, increasing the critical migration rate. The rate of migration can slow as the planets approach the inner edge of the disk due to the decrease in surface density \citep{Kley2014, Kley2015}. This slowing makes the likelihood of scattering more probable since it increases the time of interaction with the N:1 resonances. Additionally, if there exists a lack of circumbinary planets and especially multi-planet systems, this might be indicative of slow migration rates.

\subsection{Application to the Pluto and Charon Binary}

With a mass ratio of 0.1184, the dynamics around Pluto and its largest moon, Charon, behaves similar to the binary star systems described above. 
The binary dwarf planet has four discovered circumbinary moons: Styx, Nix, Kerberos, and Hydra. These moons are near to the 3:1, 4:1, 5:1, and 6:1 resonances with Charon and are therefore close to resonances with each other. None of these resonances are currently active according to the best estimates for the orbits and masses of the moons \citep{Showalter2015}. The very low eccentricities of all the bodies in the system makes active resonances unlikely due to the small resonance widths. \cite{Showalter2015} suggests the masses of the circumbinary moons are larger than predicted by albedo alone \citep{Brozovi2015} due to a librating three body resonant angle for higher masses. While moon-moon resonances may help constrain the mass of the moons, the location of the moons near the N:1 resonances remains unexplained.

The very low eccentricity of the moons suggests that in situ formation is likely \citep{Brozovi2015} but the collision from which the binary formed is not thought to have created enough leftover material at large enough distance to form the moons at their current location \citep{Canup2005, Canup2011}. \cite{2015BromleyKenyon} demonstrate that processes like physical collisions and collective gravitational scattering can expand the debris in the system to larger distances, facilitating formation. Resonant transport cannot explain the migration of all moons simultaneously since different N:1 resonances require different binary eccentricity for migration \citep{Cheng2014,Lithwick2008}. Resonant transport due to an eccentric Charon would also excite the eccentricity of the moons. Even outside of a resonance, the secular effects can excite the eccentricity of the moons above their current values. Formation models that predict a high initial eccentricity for the binary need to explain the current low eccentricity of the moons and overcome the likelihood of instabilities from the number of resonances in an eccentric system.

Previous efforts have used stability criteria to constrain the maximum eccentricity of Charon as well as the maximum mass of the moons \citep{Youdin2012,2017Smullen}. The instabilities described here can also limit the possible masses and formation history of the moons. The potential 5:6 Kerberos-Hydra resonance overlaps with the 6:7 resonance at $e=0.12$ for each body for the maximum mass consistent with \cite{Brozovi2015} meaning that at the current location a binary eccentricity of $e_{bin}\simeq 0.18$ is enough to cause chaos through moon-moon resonance overlap with eccentricity excitation through secular perturbations from the binary. The 5:6 Kerberos-Hydra resonance overlaps with the 11:13 resonance at $e_{Kerberos}=e_{Hydra}\simeq 0.09$ meaning that a binary eccentricity of $e_{bin}\simeq 0.14$ would cause chaos. In addition, interaction with the 5:1 and 6:1 resonances could drive the planets into PPR overlap as described in Section \ref{sec:combo}. Any formation model or estimates of the masses and eccentricities of the moons needs to take overlapping mean motion resonances into consideration as a consistency check. Taking the above instabilities into consideration, it is unlikely that moon-moon resonances were active while Charon had any significant eccentricity. Limits to the maximum eccentricity of Charon can be raised if the masses of the moons are lowered.

\subsection{Kepler 47}

As the only multi-planet circumbinary system discovered to date, Kepler 47 \citep{Orosz2019} is the most likely candidate system for the previously described instabilities. The stars have masses of $0.957 \pm 0.014 M_{\odot}$ and $0.342 \pm 0.003M_{\odot}$ and an orbital period of 7.45 days. The stars are orbited by three planets with periods of 49.41, 187.15, and 303.00 days.
Mass estimates for the planets are $< 26 M_{\earth}$, $7-43 M_{\earth}$, and $2-5 M_{\earth}$ for the inner, middle, and outer planets respectively. Both the orbits of the planets and the binary have low eccentricities, less than $0.07$ at the $+1\sigma$ level, and all orbits are nearly co-planar with mutual inclinations within 1.6 degrees of one another at the $+1\sigma$ level. \cite{Orosz2019} suggests that the circular, co-planar configuration is unlikely to have formed due to scattering events and that gentle migration is most likely. In the absence of the instabilities described here, the system is apparently long term stable \citep{KratterShannon2014}.

The inner planet is located between the 6:1 and 7:1 resonance but additional planets are more distant and therefore outside of low order mean motion resonances with the inner planet. The nearest, low order, planet-planet mean motion resonance is the 5:3 resonance between the middle and outer planets. Using the upper limits of the masses and eccentricities of the planets, the period ratio is still $>2\%$ from the resonant value, or over 5 times the resonance width away from the exact value. The low eccentricities of the bodies in the system make active mean motion resonances unlikely and thus instabilities via planet-planet mean motion resonance overlap are also unlikely. The 5:3 and 3:2 resonance between the middle and outer planet overlap at $e >0.3$ which is much larger the very small secular eccentricity variations from the near circular binary. As expected, Kepler 47 should be stable in its current configuration and not susceptible to the instabilities we have presented. This does not preclude previous epochs of instability and scattering in the system.

\section{Conclusion and Summary}

We have presented three new modes of instability in circumbinary multi-planet systems caused by interactions between mean motion resonances and secular perturbations from the binary. We calculated the conditions under which each instability should arise analytically, accounting for a variety of subtle affects on resonance locations and widths due to the unique system architecture of circumbinary systems. We validate our analytic predictions with a series of n-body integrations. Finally we predict the role of disk-driven migration in determining where these instabilities occur.

\begin{itemize}

\item The rich number of mean motion resonances in multi-planet circumbinary systems will shape the system architecture due to instabilities from overlapping mean motion resonances. The eccentricity of migrating resonant planets can grow due to secular or resonant interactions with the binary, leading to overlapping mean motion resonances. 
\item Migrating circumbinary planets around eccentric and unequal mass ratio binaries should be found preferentially in well separated, low degree resonances since those resonances overlap at larger eccentricity and are less susceptible to ejection due to excitation from the binary. 
\item The critical N:1 resonance that causes migrating resonant planets to become unstable is determined by comparing the resonant crossing time and the libration period of the N:1 resonances. The instability can be used to determine minimum migration rates for observed resonant planet pairs.
\item These instabilities limit the possible combinations of binary eccentricity and active planet-planet resonances in circumbinary systems such as limiting the maximum eccentricity of Charon to 0.14 if Kerberos and Hydra are in a 5:6 resonance.
\item Tidal evolution of the binary can eject planets at late times due to interactions with N:1 resonances. Instabilities due to tidal migration are perhaps more likely to result in instabilities due to the long timescales over which the N:1 resonances would be encountered.
\end{itemize}

\section*{Acknowledgements}

We thank Kat Volk, Rachel Smullen, Renu Malhotra, and Andrew Youdin for helpful discussions. We thank the reviewer for helpful comments. We gratefully acknowledge support from NSF AAG-1410174 and NASA 17-ATP17-0070.





\bibliography{mscbp.bbl} \bibliographystyle{mnras}




\appendix

\section{Modified Orbital Elements}
\label{sec:appen}
\subsection{Calculation}

Circumbinary planets at small distances from their host binary do not obey perfect Keplerian orbits. As a result, orbital elements derived under this assumption for positions and velocities in n-body simulations inadequately describe the orbital behavior of these planets. The deviations of orbits from simple ellipses can be particularly problematic when trying to identify resonances due to the unreliability of argument of pericenter in this case. 

Because circumbinary planets have a larger mean motion than an orbit around a single body, for a given distance from the system's barycenter, the body will have a higher velocity. This velocity excess presents as an artificial increase in the Keplerian eccentricity. The velocity increase also disrupts the Keplerian pericenter calculation since for low eccentricity orbits this extra velocity will appear as the planet being close to pericenter on a eccentric orbit.

Instead, we calculate ``geometric" elements from the state vectors in n-body integrations in Jacobi coordinates. 
The orbital elements are calculated by analyzing the distance from the barycenter and finding the minimum and maximum distances over a rolling time-frame. This rolling window looks at a smaller time frame of a larger run, finding the maximum and minimum distances for just over one orbital period. The Keplerian period, calculated from the state vectors, is used to estimate the window length. The rolling window allows us to find the minimum and maximum distances for each orbit and use these to calculate modified orbital elements. High frequency outputs are needed from the N-body simulations, typically at a rate of 100 per orbital period, in order to accurately capture the full orbit. For long runs, when computational time or space might be an issue at high frequency output, we use the ability to have periodic high frequency output in REBOUND.
The modified elements are in principal not coordinate frame specific, but work best in either Jacobi coordinates barycentric coordinates of the binary alone.

\begin{figure}
\centering
\includegraphics[scale=.25]{ea_new_c_2} 
\includegraphics[scale=.313]{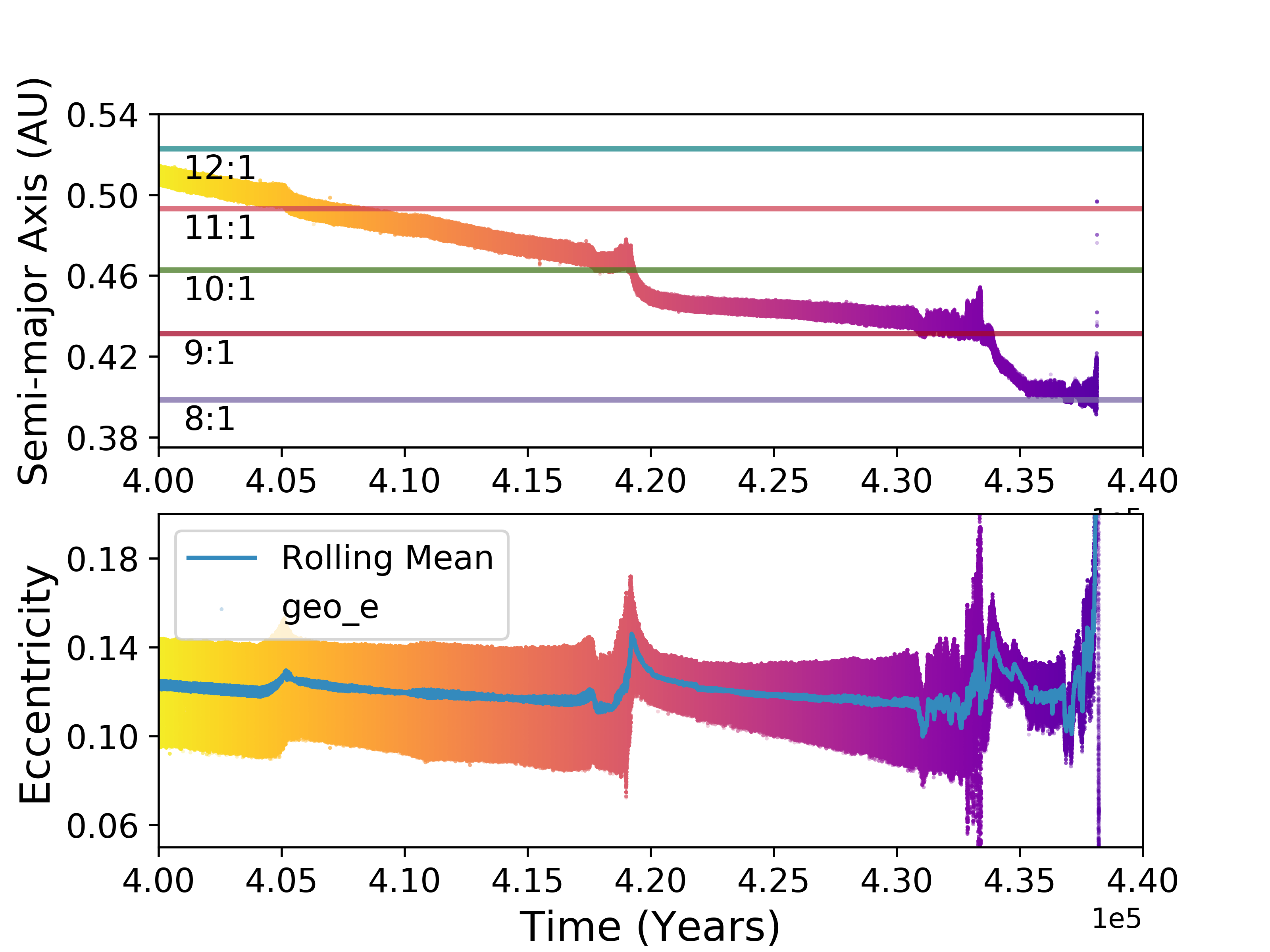}
\\
\includegraphics[scale=.25]{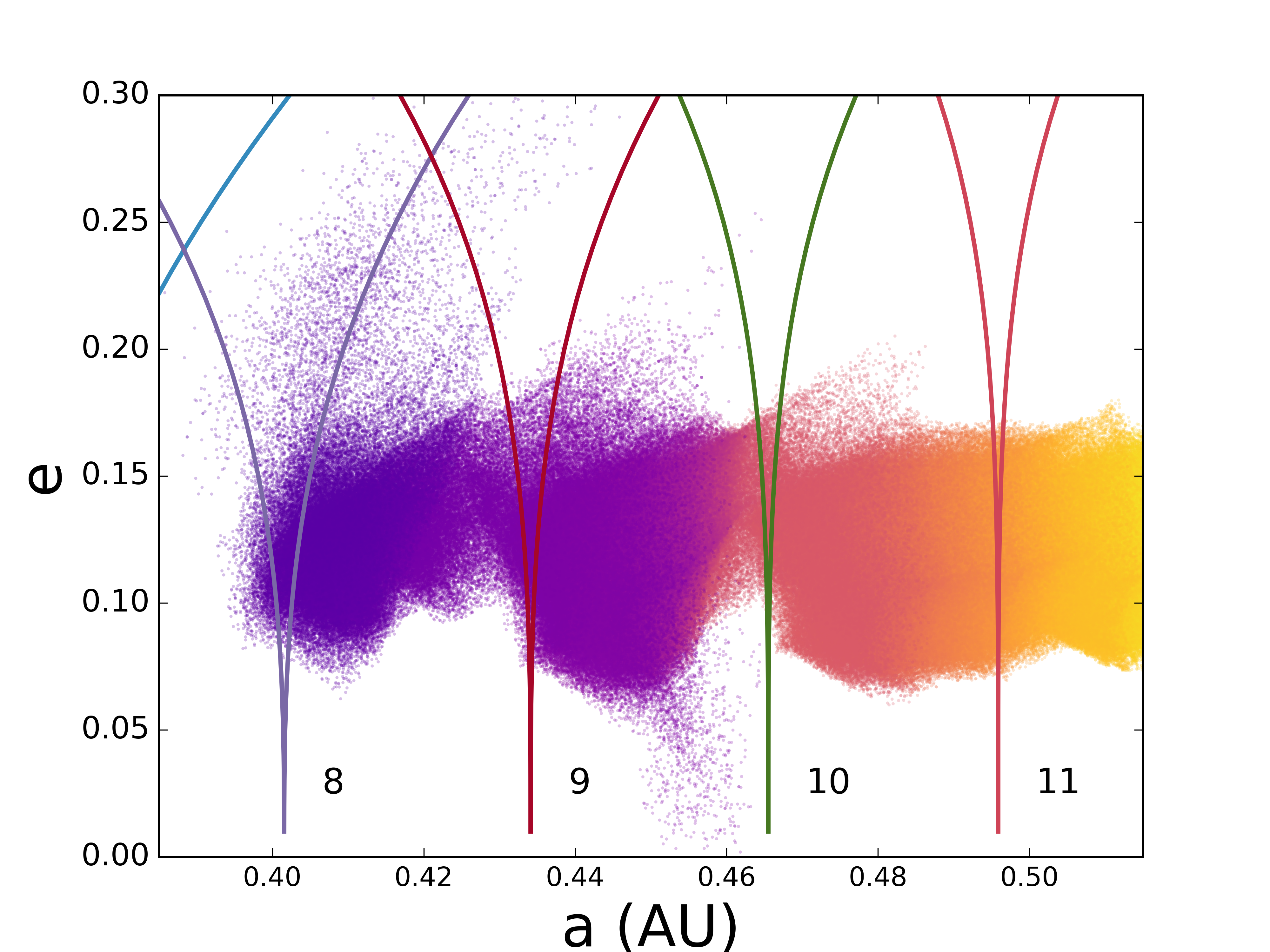} 
\includegraphics[scale=.313]{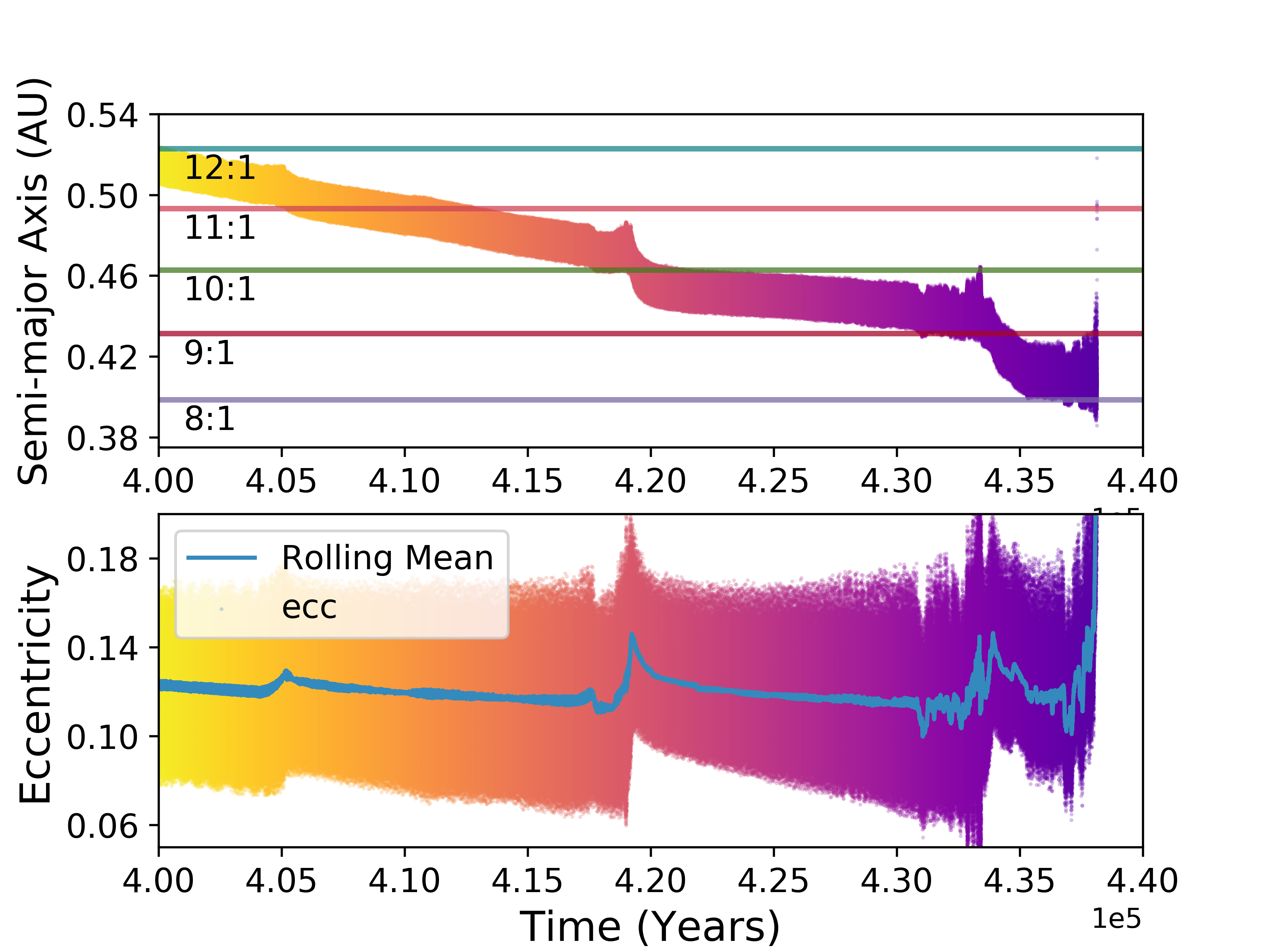}
\caption{  
Similar to figure \ref{fig:n1_num}. Top figures show the evolution of the planet using the modified orbital elements while the bottom depicts the evolution in Keplerian elements. 
}
\label{fig:kep_compare}
\end{figure}

\subsubsection{Geometric semi-major axis}
The geometric semi-major axis is calculated as
\begin{equation}
a_{geo} = (r_{max}-r_{min})/2
\label{eq:ageo}
\end{equation} 
where $r_{max}$ and $r_{min}$ are the maximum and minimum distance from barycenter for a rolling window which is as long as 2.5 times the Keplerian period.

\subsubsection{Geometric Eccentricity}
The geometric eccentricity is calculated in a similar way to the geometric semi-major axis
\begin{equation}
e_{geo} = \frac{r_{max}-r_{min}}{r_{max}+r_{min}}
\label{eq:egeo}
\end{equation}
over the same rolling window.

\subsubsection{Geometric Pericenter}

The geometric pericenter, $\omega_{geo}$, is calculated by finding the angle between the reference plane and the location of the planet at minimum barycentric distance for a rolling window which is as long as 1.25 times the Keplerian period. We then linearly interpolate to find the pericenter angle between pericenter passages.

\subsubsection{Mean Anomaly}

The mean anomaly is calculated by linearly interpolating between pericenters.


\bsp	
\label{lastpage}
\end{document}